\documentclass[12pt]{iopart}

\usepackage[T1]{fontenc}

\usepackage{amsmath1,amsthm,amssymb}

\usepackage[dvips]{graphicx}
\usepackage{colordvi}

\def\R{\mathbb R}
\def\N{\mathbb N}
\def\C{\mathbb C}

\def\rank{\mathrm{rank}}

\newtheorem{lem}{Lemma}[section]
\newtheorem{prop}[lem]{Proposition}
\newtheorem{thm}[lem]{Theorem}
\newtheorem{coro}[lem]{Corollary}
\newtheorem{obs}[lem]{Observation}

\theoremstyle{definition}

\newtheorem{pozn}[lem]{Remark}

\def\pf{\begin{proof}}
\def\pfk{\end{proof}}

\renewcommand{\d}{{\mathrm d}}
\renewcommand{\i}{{\mathrm i}}
\def\e{\mathrm e}

\def\O{\mathcal O}

\def\cotg{{\mathrm{cotg\,}}}
\def\arctg{{\mathrm{arctg\,}}}

\begin{document}

\title[High-energy asymptotics of the spectrum of a square lattice]
{High-energy asymptotics of the spectrum of a periodic square-lattice quantum graph}

\author{Pavel Exner}
\address{Doppler Institute for Mathematical Physics and Applied
Mathematics, \\ Czech Technical University, B\v{r}ehov\'{a} 7,
11519 Prague, \\ and  Nuclear Physics Institute ASCR, 25068
\v{R}e\v{z} near Prague, Czechia} \ead{exner@ujf.cas.cz}

\author{Ond\v{r}ej Turek}
\address{Laboratory of Physics, Kochi University of Technology \\
Tosa Yamada, Kochi 782-8502, Japan} \ead{ondrej.turek@kochi-tech.ac.jp}

\begin{abstract}
We investigate a periodic quantum graph in form of a square
lattice with a general self-adjoint coupling at the vertices. We
analyze the spectrum, in particular, its high-energy behaviour.
Depending on the coupling type, bands and gaps have different
asymptotics. Bands may be flat even if the edges are coupled, and
non-flat band widths may behave like $\O(n^j),\, j=1,0,-1,-2,-3$, as
the band index $n\to\infty$. The gaps may be of asymptotically
constant width or linearly growing with the latter case being
generic.
\end{abstract}

\section{Introduction} \label{s: intro}

Origins of the quantum graph concept can be traced back to Linus
Pauling's considerations about the structure of organic molecules in
1930's, however, it was roughly the last two decades when the
subject became very popular and its true richness has been
revealed. The reason for that was not only the fact that quantum
graphs represent a suitable model of numerous systems prepared by
solid state physicists. Equally important was the inherent
structure of such models which allowed us to study effects
uncommon in the ``usual'' quantum mechanics coming from the
nontrivial topological structure of graph structures as well as
from the fact that they mix features corresponding to different
dimensionalities. The bibliography concerning quantum graphs is
nowadays extensive indeed and we restrict ourselves to quoting the
proceedings volume of a recent topical programme at Isaac Newton
Institute and references therein \cite{AGA}.

One often studied class are periodic graphs. Their spectrum has
predictably a band structure, however, in distinction to the usual
Schr\"odinger operators they can exhibit under particular
geometric conditions also infinitely degenerate eigenvalues
manifesting invalidity of the unique continuation principle
\cite{Ku2}. The aim of the present paper is to investigate
the spectrum in a particular example of periodic graphs, namely
two-dimensional square lattices with a general self-adjoint
coupling at the vertices. This generalizes earlier work in
which lattices with $\delta$ and $\delta'_\mathrm{s}$ couplings
were studied \cite{E1, EG}.

A motivation for the present extension comes from different
sources. First of all, the general vertex coupling became more
interesting after several recent results --- cf.~\cite{KZ, EP,
CET} and references therein --- showing how it can be approximated
by graphs or network systems with suitably scaled potentials
illustrating thus that it is not just a mathematical object but
also something which can be, in an approximative sense at least,
realized physically. Secondly, already the mentioned examples of
$\delta$ and $\delta'_\mathrm{s}$ couplings demonstrated that
different couplings yield different asymptotical behaviour of
spectral bands and gaps at high energies.

These asymptotics can have interesting dynamical consequences.
Recall that in the one-dimensional analogue of the present
problem, in the generalized Kronig-Penney model, there are three
types of asymptotic behaviour \cite{EGr, CS} and that the one
corresponding to the $\delta'$ interaction for which spectral gaps
are dominating exhibit absence of transport in the Wannier-Stark
situation when an electric field is applied \cite{AEL, E2, MS,
ADE}; we stress that the solution of the Wannier-Stark problem in the
other two cases mentioned is still an open question. With these
facts in mind it is natural to ask how many types of asymptotic
behaviour a two dimensional square lattice can show and what they
look like. Our approach to this problem is based on the use of the
so-called $ST$-form of boundary condition introduced in
\cite{CET}. This will allow us to express the band and gap widths
by rather simple expressions involving the coupling parameters,
from which it is easy to determine how the spectral behaviour is
governed by the particular boundary conditions.

We will show that the high-energy behaviour has more types than in
the one-dimensional situation. In fact, the problem has sixteen
parameters and offers a zoology of solutions. The scope of this
issue does not allow us to present a complete classification but
we will list all the ``generic'' cases with respect to the rank of
the matrix $B$ in the condition \eqref{bc:KS} below and single out
cases of particular interest. The two-dimensional character of the
problem has two main consequences. The first one is a possible
occurrence of flat bands, or infinitely degenerate eigenvalues;
this is connected with the invalidity of the principle of unique
continuation on graphs of nontrivial topology mentioned above. The
second remarkable feature is the existence of couplings for which
the spectral bands are shrinking as $n^{-1}$, $n^{-2}$, or
$n^{-3}$ with respect to the band index $n$. This effect again has
no analogue in the one-dimensional situation. The most ``generic''
situation, however, is the one known from \cite{E1, EG} when the
bands are asymptotically of constant widths and gaps are linearly
growing.

\section{Preliminaries and main result} \label{s: prelim}

\subsection{Square lattice with a general vertex coupling}

As we have said we will consider a square lattice graph. To be
concrete, the vertices of $\Gamma$ are $\{(ma,na):\:
m,n\in\mathbb{Z}\}$ for a fixed $a>0$ and the edges are segments
of length $a$ connecting points differing by one in one of the two
indices. The state Hilbert space is the orthogonal sum of the
$L^2$ spaces on the edges and the Hamiltonian acts as
$-\frac{\mathrm{d}^2}{\mathrm{d}x^2}$ on each of them, with the
domain consisting of the corresponding $W^{2,2}$ functions.

It is well known that in order to get a self-adjoint operator one
has to impose boundary conditions at graph vertices which couple
the vectors $\Psi(0)$ and $\Psi'(0)$ of the boundary values --- we
choose the variables at all the adjacent edges so that they start
at the vertex. The standard form of these conditions is
 \begin{equation} \label{bc:KS}
A\Psi(0) + B\Psi'(0)=0\,,
 \end{equation}
where $A,\,B$ are matrices such that $(A|B)$ has maximum rank and
$AB^*$ is self-adjoint \cite{KS1}. Being interested in the
periodic situation we naturally assume that the matrices $A,\,B$
are the same at each vertex of the lattice. A drawback of the
conditions \eqref{bc:KS} is that the matrix pair is not unique.
There are various ways to mend this problem. One is to rewrite
\eqref{bc:KS} in the form
 \begin{equation} \label{bc:Ha}
(U-I)\Psi(0) +i(U+I)\Psi'(0)=0\,,
 \end{equation}
where $U$ is a unitary matrix. In the quantum graph context these
conditions have been proposed in \cite{Ha, KS2}, however, they
were known earlier in the general theory of boundary forms
\cite{GG}. It is clear that a distinguished role is played by
subspaces of the boundary space values referring to eigenvalues
$\mp 1$ of $U$. An alternative way \cite{Ku1} to write the
conditions is by means of the corresponding orthogonal projection
$P$ and its complement $Q:=I-P$: there is a self-adjoint operator
$L$ in $Q\C^n$ such that
 \begin{equation} \label{bc:Ku}
P\Psi(0)=0\,, \quad Q\Psi'(0)+LQ\Psi(0)=0\,.
 \end{equation}
Here we are going to use yet another version introduced in
\cite{CET} as the \emph{$ST$-form},
\begin{equation}\label{bc:ST}
\left(\begin{array}{cc}
I^{(m)} & T \\
0 & 0
\end{array}\right)\Psi'(0)=
\left(\begin{array}{cc}
S & 0 \\
-T^* & I^{(n-m)}
\end{array}\right)\Psi(0)
\end{equation}
using matrices $T\in\C^{m,n-m}$ and a self-adjoint $S\in\C^{m,m}$,
where $n$ is the dimension of the boundary value space (which will
be four in our case) and $m=n-\dim P\in \{0,\dots,n\}$.

\subsection{The main result}

Let $H_{S,T}$ is the quantum graph Hamiltonian described above.
Our results about its spectrum can be summarized as follows:

\begin{thm}
(a) The spectrum of $H_{S,T}$ consists of absolutely continuous
spectral bands and infinitely degenerate eigenvalues. Its negative
part consists of at most four bands. \\ [.2em]
(b) The high-energy asymptotic behaviour of spectral bands
and gaps as a function of the band index $n$ includes the
following classes:
 \begin{itemize}
 \item flat bands, i.e. infinitely degenerate point spectrum,
 \item bands behaving like $\O(n^j),\, j=1,0,-1,-2,-3$, as $n\to\infty$,
 \item gaps behaving like $\O(n^j),\, j=1,0$, as $n\to\infty$.
 \end{itemize}
Depending on the vertex coupling \eqref{bc:KS} the high-energy
asymptotics of the spectrum may be a combination of the above
listed types.
\end{thm}

The rest of the paper is devoted to demonstrating of these claims.
We will do that by analyzing spectrum of the fiber operator coming
from the Bloch-Floquet analysis, discussing subsequently
situations corresponding to different values of rank $m$ of the
matrix $B$ in the boundary conditions (\ref{bc:KS}). While this
procedure serves best our aim, it is not very illustrative from of
the point of view of particular type of edge coupling. Apart from
the trivial case $m=0$ when the lattice decomposes into separate
edges with Dirichlet conditions, each of the other values of $m$
cover several subcases with very different couplings and spectral
behaviours. They would deserve a separate discussion which we
cannot present here due to volume restrictions and we postpone it
to another publication; we limit ourselves to several general
statements:

\begin{itemize}

\item The generic situation corresponds to $m=4$ with all the
edges coupled and spectral gaps growing linearly with the band
index,

\item each case $m=1,2,3,4$ covers a situations with flat bands
corresponding to lattice decoupling into separated edges, or pairs
of edges,

\item the lattice can separate into ``one-dimensional'' subsets
describing generalized Kronig-Penney models on lines or zigzag
curves, or to ``combs'',

\item from the spectral point of view the case $m=3$ is the
richest, including situations with a powerlike shrinking of
spectral bands that occurs for the graph decomposed into ``combs''.

\end{itemize}

\section{Bloch-Floquet analysis}

\begin{figure}[h]
\begin{center}
\hspace*{-6em}\includegraphics[angle=0,width=0.3\textwidth]{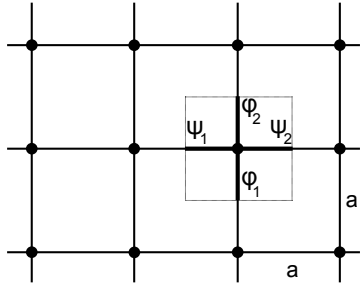}
\caption{\emph{A periodic two-dimensional network}}\label{Mrizka}
\end{center}
\end{figure}

Since our graph is $a$-periodic w.r.t. shifts in both directions,
we are able to employ the Bloch-Floquet decomposition. The
elementary cell is depicted in Fig.~\ref{Mrizka}, together with
the notation
of the wave function components on the edges.

The analysis follows the same pattern as in \cite{E1, EG} but the
general vertex coupling makes it substantially more complicated.
The fiber operator corresponding to fixed values of Floquet
parameters (quasimomentum components) has a purely discrete
spectrum and the number of its negative eigenvalues is at most
four. The last claim follows from general principles \cite[Sec.
8.3]{We} and from a comparison to the square graph Hamiltonian
with Kirchhoff coupling at the vertices the spectrum of which
equals by \cite{E1} $\mathbb{R}^+$; it is enough to notice that
the Floquet component of this operator and that of a general
$H_{S,T}$ have a common symmetric restriction with deficiency
indices $(n,n),\; n\le 4$. Note also that the bands may overlap in
general; an example can be constructed using boundary conditions
which separate motion in the two directions leading to two
families of generalized Kronig-Penney models.

Our main interest concerns the positive part of the spectrum, and
as usual we are thus going to investigate solutions of the
corresponding stationary Schr\"odinger equation with energy
$E=k^2,\;k>0$. It is obvious that they are at each edge linear
combinations of the functions $\e^{\i kx}$ and $\e^{-\i kx}$,
specifically
\begin{equation}\label{vlnfce}
\begin{split}
\psi_1(x)&=C_1^+\e^{\i k x}+C_1^-\e^{-\i k x},\quad x\in[-a/2,0]\\
\psi_2(x)&=C_2^+\e^{\i k x}+C_2^-\e^{-\i k x},\quad x\in[0,a/2]\\
\varphi_1(x)&=D_1^+\e^{\i k x}+D_1^-\e^{-\i k x},\quad x\in[-a/2,0]\\
\varphi_2(x)&=D_2^+\e^{\i k x}+D_2^-\e^{-\i k x},\quad x\in[0,a/2]
\end{split}
\end{equation}
By assumption, they have to satisfy the boundary conditions at the
vertex, i.e. it holds
\begin{equation}\label{vazba}
A\left(\begin{array}{c}\psi_1(0)\\ \psi_2(0)\\ \varphi_1(0)\\
\varphi_2(0)\end{array}\right)+B\left(\begin{array}{c}-\psi_1'(0)\\
\psi_2'(0)\\ -\varphi_1'(0)\\ \varphi_2'(0)\end{array}\right)=0\,.
\end{equation}
In addition to that, of course, they have to satisfy Bloch-Floquet
conditions,
\begin{equation}\label{Floquet}
\begin{split}
\psi_2(a/2)=\e^{\i\theta_1}\psi_1(-a/2)\,&\qquad\qquad
\varphi_2(a/2)=\e^{\i\theta_2}\varphi_1(-a/2)\,,\\
\psi_2'(a/2)=\e^{\i\theta_1}\psi_1'(-a/2)\,&\qquad\qquad
\varphi_2'(a/2)=\e^{\i\theta_2}\varphi_1'(-a/2)\,,
\end{split}
\end{equation}
for fixed values of the quasimomentum components
$\theta_1,\theta_2\in(-\pi,\pi]$. Substituting~\eqref{vlnfce}
into~\eqref{Floquet} allows us to express the variables $C_2^{\pm}$
and $D_2^{\pm}$ in terms of $C_1^{\pm}$ and $D_1^{\pm}$,
\begin{equation}\label{elim}
C_2^\pm =C_1^\pm\cdot\e^{\i(\theta_1\mp ak)}\,, \quad D_2^\pm
=D_1^\pm\cdot\e^{\i(\theta_2\mp ak)}\,.
\end{equation}
Using these relations we eliminate $C_2^{\pm}$ and $D_2^{\pm}$
from~\eqref{vlnfce}, after that we substitute~\eqref{vlnfce}
into~\eqref{vazba}. Simple manipulations then yield the following
condition,
\begin{equation}\label{matice}
\left[(AM+\i kBN)D\right]
\left(\begin{array}{c}C_1^+\\C_1^-\\D_1^+\\D_1^-\end{array}\right)=0\,,
\end{equation}
where $D :=\mathrm{diag}\left(\e^{\frac{\i}{2}(\theta_1-ak)},
\e^{\frac{\i}{2}(\theta_1+ak)},\e^{\frac{\i}{2}(\theta_2-ak)},
\e^{\frac{\i}{2}(\theta_2+ak)}\right)$ and the matrices $M,\,N$
are given by
$$
M:=\left(\begin{array}{cccc}
\e^{-\frac{\i}{2}(\theta_1-ak)}&\e^{-\frac{\i}{2}(\theta_1+ak)}&0&0\\
\e^{\frac{\i}{2}(\theta_1-ak)}&\e^{\frac{\i}{2}(\theta_1+ak)}&0&0\\
0&0&\e^{-\frac{\i}{2}(\theta_2-ak)}&\e^{-\frac{\i}{2}(\theta_2+ak)}\\
0&0&\e^{\frac{\i}{2}(\theta_2-ak)}&\e^{\frac{\i}{2}(\theta_2+ak)}
\end{array}
\right)\,,
$$
$$
N:=\left(\begin{array}{cccc}
-\e^{-\frac{\i}{2}(\theta_1-ak)}&\e^{-\frac{\i}{2}(\theta_1+ak)}&0&0\\
\e^{\frac{\i}{2}(\theta_1-ak)}&-\e^{\frac{\i}{2}(\theta_1+ak)}&0&0\\
0&0&-\e^{-\frac{\i}{2}(\theta_2-ak)}&\e^{-\frac{\i}{2}(\theta_2+ak)}\\
0&0&\e^{\frac{\i}{2}(\theta_2-ak)}&-\e^{\frac{\i}{2}(\theta_2+ak)}
\end{array}
\right)\,.
$$
It follows from~\eqref{elim} that the functions~\eqref{vlnfce}
correspond to a nonzero solution \emph{iff} $\left(C_1^+,C_1^-,
D_1^+,D_1^-\right)$ is a nonzero vector. Consequently, a number
$k^2$ belongs to the spectrum of the Hamiltonian if and only
if~\eqref{matice} has a non-trivial solution for some pair
$(\theta_1,\theta_2)$, in other words, if there are
$\theta_1,\,\theta_2$ such that $\det\left[(AM+\i kBN)D\right]=0$
which can be simplified to
\begin{equation}\label{ObecSp}
\det\left(AM+\i kBN\right)=0\,.
\end{equation}
Our aim is to analyze the spectral asymptotics in dependence on
the coupling type. Since four edges join at each vertex, the
problem has 16 real parameters. We will take them into account
through the $ST$ form \eqref{bc:ST} of the boundary conditions,
i.e. we set
$$
-A=
\left(\begin{array}{cc}
S & 0 \\
-T^* & I^{(4-m)}
\end{array}\right)\,,\quad
B=\left(\begin{array}{cc}
I^{(m)} & T \\
0 & 0
\end{array}\right)\,,
$$
where $m:=\rank(B)$. Obviously, each value of $m$ has to be discussed separately.

\section{The case of $m=0$, or Dirichlet decoupled edges}

Consider first the simplest situation when $B=0,\: A=-I$. The
$ST$-form of boundary conditions is then obviously invariant with
respect to the edge labelling and the spectral condition
\eqref{ObecSp} acquires the form $\det(-M)=0$. Since
$\det(-M)=\det(M)=-4\sin^2 ak$, this requires $\sin ak=0$, hence
the spectrum consists of infinitely degenerate eigenvalues,
$$
\sigma(H)=\left\{\left(\left.\frac{n\pi}{a}\right)^2\,\right|\,n\in\N\right\}\,.
$$

\section{The case of  $m=1$}

The admissible couplings form a seven-parameter family corresponding to the choice
$$
B=\left(\begin{array}{cccc}
1 & t_1 & t_2 & t_3 \\
0 & 0 & 0 & 0 \\
0 & 0 & 0 & 0 \\
0 & 0 & 0 & 0
\end{array}\right)\,,\quad
A=
-\left(\begin{array}{cccc}
s & 0 & 0 & 0 \\
-\overline{t_1} & 1 & 0 & 0 \\
-\overline{t_2} & 0 & 1 & 0 \\
-\overline{t_3} & 0 & 0 & 1
\end{array}\right)
$$
with $s\in\R$ and $t_j\in\C$. Note that while in general the
boundary conditions in the $ST$-form are not invariant with
respect to the edge renumbering, we can choose without loss of
generality the ``privileged one'' corresponding to the first
component. A direct calculation of the determinant in
\eqref{ObecSp} leads to the spectral condition
$$
-4k\sin ak\left[(1+|t_1|^2+|t_2|^2+|t_3|^2)\cos ak
-2\Re(t_1\e^{\i\theta_1}+\overline{t_2}t_3\e^{\i\theta_2})\right]-4s\,\sin^2
ak=0\,.
$$
We see that the condition is solved again by $\frac{n\pi}{a}$ for
$n\in\N$. In addition to that, there are solutions coming from the
equation
\begin{equation}\label{m1.1}
(1+|t_1|^2+|t_2|^2+|t_3|^2)\cos
ak-2\Re(t_1\e^{\i\theta_1}+\overline{t_2}t_3\e^{\i\theta_2})
=-\frac{1}{k}\, s\,\sin ak\,.
\end{equation}
Recall that $k^2>0$ is in the spectrum if there are $\theta_1,
\theta_2\in(-\pi,\pi]$ which together with $k$ satisfy
\eqref{m1.1}. It is convenient to rewrite the range of the
expression $\Re(t_1 \e^{\i\theta_1} +\overline{t_2}t_3
\e^{\i\theta_2})$ using a single parameter as follows:
\begin{obs}\label{m1.o1}
It holds
$$
\left\{\left.\Re(t_1\e^{\i\theta_1}
+\overline{t_2}t_3\e^{\i\theta_2})\,\right|\,\theta_1,\theta_2\in(-\pi,\pi]\right\}=
\left\{\left.(|t_1|+|t_2|\cdot|t_3|)\cos\vartheta\,\right|\,\vartheta\in(-\pi,\pi]\right\}\,.
$$
\end{obs}

\medskip

\noindent The spectral condition \eqref{m1.1} then yields the
following requirement:
\begin{equation}\label{m1.2}
(1+|t_1|^2+|t_2|^2+|t_3|^2)\cos
ak-2(|t_1|+|t_2|\cdot|t_3|)\cos\vartheta =-\frac{1}{k}\,s\,\sin
ak\,.
\end{equation}
This conditions has various types of solutions in dependence on
the parameter values. Before discussing them, let us mention
another useful fact.
\begin{obs}\label{m1.o2}
It holds $2(|t_1|+|t_2|\cdot|t_3|)\leq 1+|t_1|^2+|t_2|^2+|t_3|^2$
and the equality occurs if and only if $|t_1|=1\:\wedge\:
|t_2|=|t_3|$.
\end{obs}

\subsection{Point spectrum}

If the expression $|t_1|+|t_2|\cdot|t_3|$ vanishes, i.e. if
$t_1=0$ and $t_2=0\:\vee\: t_3=0\:$ (we may suppose without loss of
generality that $t_3=0$), then \eqref{m1.1} becomes
\begin{equation}\label{m1.3}
\cotg ak=-\frac{1}{k}\,\frac{s}{1+|t_2|^2}\,.
\end{equation}
The \emph{rhs} being $\mathcal{O}(k^{-1})$ as $k\to\infty$ the
solutions are obviously close to the numbers $(-\frac{1}{2}+n)
\frac{\pi}{a}$ at high energies. Writing them as
$$
k=\left(-\frac{1}{2}+n\right)\frac{\pi}{a}+\delta\,;
$$
we get $\cotg ak=a\delta+\O(\delta^3)$ and $\frac{1}{k}
=\frac{a}{n\pi}+\O(n^{-2})$. A substitution into \eqref{m1.3} and
a few manipulations give $\delta=\frac{1}{n\pi} \cdot\frac{-s}
{1+|t_2|^2}+\O\left(n^{-2}\right)$, hence this solution represents
the spectral points $k^2=k_n^2$ behaving like
$$
k^2=\left[\left(-\frac{1}{2}+n\right)\frac{\pi}{a}\right]^2
+\frac{2}{a}\cdot\frac{-s}{1+|t_2|^2}+\O(n^{-1})
$$
for $n\to\infty$. These solutions are independent of the quasimomentum
and give rise to flat spectral bands; it is not difficult to see
that the corresponding eigenfunctions can be chosen compactly
supported.

Let us remark that the boundary conditions studied above mean that
the Hamiltonian decouples into a countable direct sum of operators
supported on two edges of the graph, or on individual edges if
$T=0$ and $s=0$, and therefore it is not surprising that the
spectrum is pure point.

\subsection{Linearly growing spectral bands and gaps}\label{m1.LinGr}

If $|t_1|+|t_2|\cdot|t_3|\neq0$, we can divide by it and express
thus $\cos\vartheta$ from \eqref{m1.2}. This yields the spectral
condition in the form
\begin{equation}\label{m1.4}
\left|\frac{1+|t_1|^2+|t_2|^2+|t_3|^2}{2(|t_1|+|t_2|\cdot|t_3|)}\cos ak
+\frac{1}{k}\cdot\frac{s}{2(|t_1|+|t_2|\cdot|t_3|)}\sin ak\right|\leq1\,.
\end{equation}
By Observation \ref{m1.o2}, the coefficient of $\cos ak$ at the
\emph{lhs} cannot be smaller than one.

In the rest of this section, we focus on the case
$|t_1|\neq1\:\vee\: |t_2|\neq |t_3|$; the remaining situation will
be treated in Section~\ref{m1.s.acg}. Since the coefficient of
$\sin ak$ is $\O(k^{-1})$, it is evident that \eqref{m1.4} can be
asymptotically satisfied only away from the points where $|\cos
ak|=1$, in other words, the spectral bands are neigbourhoods of
the points
$\left[\left(-\frac{1}{2}+n\right)\frac{\pi}{a}\right]^2$. Let us
set $k=\left(-\frac{1}{2}+n\right)\frac{\pi}{a}+d$ and find the
range of $d$. We see that $\cos ak=\sin ad$ and $\frac 1k$ of the
$n$-th band solution is $\O(n^{-1})$, thus \eqref{m1.4} can be
rewritten as
$$
|\sin ad|\leq\frac{2(|t_1|+|t_2|\cdot|t_3|)}
{1+|t_1|^2+|t_2|^2+|t_3|^2}+\O(n^{-1})\,.
$$
Hence $ad\leq\Delta:=\arcsin\frac{2(|t_1|+|t_2|\cdot|t_3|)}{1+|t_1|^2+|t_2|^2+|t_3|^2}$;
note that $\Delta\in\left(0,\frac{\pi}{2}\right)$ in view of the assumption and
Observation \ref{m1.o2}. Then the spectral bands behave like
$$
\left(\left[\left(-\frac{1}{2}+n\right)\frac{\pi}{a}
-\frac{\Delta}{a}+\O\left(\frac{1}{n}\right)\right]^2,
\left[\left(-\frac{1}{2}+n\right)\frac{\pi}{a}
+\frac{\Delta}{a}+\O(n^{-1})\right]^2\right)\,,
$$
which can be rewritten as
$$
\left(\left(-\frac{1}{2}+n\right)^2\frac{\pi^2}{a^2}
-2\,\frac{n\pi}{a^2}\Delta+\O(1),\left(-\frac{1}{2}+n\right)^2
\frac{\pi^2}{a^2}+2\,\frac{n\pi}{a^2}\Delta+\O(1)\right)
$$
in the high-energy limit, $n\to\infty$. In other words, both bands
and gaps are asymptotically linearly growing with the band index.

\subsection{Asymptotically constant spectral gaps}\label{m1.s.acg}

If $|t_1|=1$ and $|t_2|=|t_3|\neq0$, the spectral condition
\eqref{m1.4} has to be treated differently being now of the form
\begin{equation}\label{m1.5}
\left|\cos ak+\frac{1}{k}\cdot\frac{s}{2(1+|t_2|^2)}\sin ak\right|\leq1\,.
\end{equation}
Let us suppose that $s\neq0$ --- the case $s=0$ is special and
will be discussed below in Section~\ref{m1.s.ng}. Since $|\cos
ak|\leq1$ and the coefficient at $\sin ak$ is small in modulus for
large values of $k$, we see that the condition \eqref{m1.5} is
violated only in a small one-sided neighbourhood of the points
where $|\cos ak|=1$. To describe the corresponding \emph{gaps}, we
set $k=\frac{n\pi}{a}+\delta$, then we have
\begin{gather*}
\cos ak=(-1)^n\cdot\left(1-\frac{(a\delta)^2}{2}+\O\left(\delta^4\right)\right)\,,\\
\sin ak=(-1)^n\cdot a\delta+\O\left(\delta^3\right)\,,\\
\frac{1}{k}=\frac{a}{n\pi}+\O\left(\frac{\delta}{n^2}\right)\,.
\end{gather*}
Substituting from here into the \emph{negated}
condition \eqref{m1.5} we obtain the gap condition,
$$
\left|1-\frac{(a\delta)^2}{2}
+\frac{a^2}{n\pi}\frac{s}{2(1+|t_2|^2)}\,\delta+
\O\left(\delta^4\right)+\O\left(\frac{\delta^3}{n}\right)
+\O\left(\frac{\delta^2}{n^2}\right)\right|>1\,,
$$
which is, in dependence of the sign of $s$, solved by
$$
\delta\in\left\{\begin{array}{ll}
\left(\O\left(\frac{1}{n^2}\right),
\frac{1}{n\pi}\cdot\frac{s}{1+|t_2|^2}
+\O(n^{-2})\right) & \text{for}\quad s>0\,, \vspace*{1mm}\\
\left(\frac{1}{n\pi}\cdot\frac{s}{1+|t_2|^2}
+\O\left(\frac{1}{n^2}\right),\O(n^{-2})\right) & \text{for}\quad
s<0\,.
\end{array}\right.
$$
Consequently, the gap boundaries are
$$
\left(\frac{n\pi}{a}\right)^2+\O(n^{-1}) \qquad \text{and} \qquad
\left(\frac{n\pi}{a}\right)^2+\frac{2}{a}\cdot\frac{s}{1+|t_2|^2}+\O(n^{-1})\,,
$$
in other words, the gap widths are asymptotically constant and
spectral bands grow linearly w.r.t. the band number $n$. Note that this
case includes lattices with a \emph{nontrivial $\delta$ coupling}
discussed in \cite{E1, EG}.

\subsection{No gaps, spectrum on the nonnegative halfline}\label{m1.s.ng}

It remains to discuss the case with  $|t_1|=1$, $|t_2|=|t_3|\neq0$
and $s=0$, when the spectral condition \eqref{m1.5} simplifies to
$\left|\cos ak\right|\leq1$ which is obviously satisfied for all
$k>0$. Moreover, one checks directly that $0\in\sigma(H)$, and
putting $k=\i\kappa$ we find $\sigma(H)\cap(-\infty,0)=\emptyset$,
hence $\sigma(H)=[0,+\infty)$. Referring again to \cite{E1, EG} we
note that this includes the case of a lattice with \emph{Kirchhoff
coupling}.


\section{The case of  $m=2$}

The reader has noted already that the boundary conditions in the
$ST$-form are not invariant with respect to the edge renumbering.
Neglecting trivial lattice replacement corresponding to rotations
and mirror images, we must distinguish two situations here:
\begin{itemize}
\item[(i)] Linearly independent columns of $B$ are associated with
parallel edges; without loss of generality we may suppose they are
the ``horizontal'' ones. Then we apply the conditions in $ST$-form
directly,
$$
B=\left(\begin{array}{cccc}
1 & 0 & t_{11} & t_{12} \\
0 & 1 & t_{21} & t_{22}\\
0 & 0 & 0 & 0 \\
0 & 0 & 0 & 0
\end{array}\right)\,,\quad
A=
-\left(\begin{array}{cccc}
s_{11} & s_{12} & 0 & 0 \\
\overline{s_{12}} & s_{22} & 0 & 0 \\
-\overline{t_{11}} & -\overline{t_{21}} & 1 & 0 \\
-\overline{t_{12}} & -\overline{t_{22}} & 0 & 1
\end{array}\right)\,,
$$
where $s_{11},\, s_{22}$ are real and the other matrix entries are
complex.
\item[(ii)] Linearly independent columns of $B$ can correspond
also to mutually orthogonal edges, say, the left ``horizontal''
and the lower ``vertical''. Then we use the conditions in a
permuted form, with second and third row of $\Psi(0)$ and
$\Psi'(0)$ interchanged. Since it is convenient to keep the
entries order in these vectors, we interchange instead the second
and the third column of the matrices $A,\,B$,
$$
B=\left(\begin{array}{cccc}
1 & t_{11} & 0 & t_{12} \\
0 & t_{21} & 1 & t_{22}\\
0 & 0 & 0 & 0 \\
0 & 0 & 0 & 0
\end{array}\right)\,,\quad
A=
-\left(\begin{array}{cccc}
s_{11} & 0 & s_{12} & 0 \\
\overline{s_{12}} & 0 & s_{22} & 0 \\
-\overline{t_{11}} & 1 & -\overline{t_{21}} & 0 \\
-\overline{t_{12}} & 0 & -\overline{t_{22}} & 1
\end{array}\right)\,.
$$
\end{itemize}
However, since the spectral analysis of situations (i) and (ii)
can be done in a very similar way, and moreover, also the
structure of the spectral bands is essentially the same, we
perform the analysis of the case (i) only.

The determinant in \eqref{ObecSp} leads to the spectral condition
which can be written as
$$
V_2\cdot k^2+V_1\cdot k+V_0=0\,,
$$
where $V_2$, $V_1$ and $V_0$ are expressions depending on $ak$, on
the entries of $S$, $T$, and on the quasimomentum components
$\theta_1,\theta_2$. By a direct computation we get
\begin{align*}
V_2&=-4\cos^2 ak\left[|t_{11}|^2+|t_{22}|^2+|t_{12}|^2+|t_{21}|^2\right]+4\sin^2 ak\left[1+\left|t_{11}t_{22}-t_{12}t_{21}\right|^2\right]\\
&+8\cos ak\left[-\Re\left((t_{11}\overline{t_{21}}+t_{12}\overline{t_{22}})\e^{\i\theta_1}\right)+
\Re\left((t_{22}\overline{t_{21}}+\overline{t_{11}}t_{12})\e^{\i\theta_2}\right)\right]\\
&+8\Re\left[t_{11}\overline{t_{22}}\e^{\i(\theta_1-\theta_2)}\right]
+8\Re\left[t_{12}\overline{t_{21}}\e^{\i(\theta_1+\theta_2)}\right]\,, \\[.5em]
V_1&=4\sin ak\left[-\cos ak\left(s_{11}(1+|t_{21}|^2+|t_{22}|^2)+s_{22}(1+|t_{11}|^2+|t_{12}|^2)\right.\right.\\
&\left.-2\Re\left(s_{12}(\overline{t_{11}}t_{21}+\overline{t_{12}}t_{22})\right)\right)\\
&\left.-2\Re\left(s_{12}\e^{\i\theta_1}\right)
+2\Re\left((s_{11}\overline{t_{21}}t_{22}+s_{22}\overline{t_{11}}t_{12}
-s_{12}\overline{t_{11}}t_{22}-s_{12}\overline{t_{12}}t_{21})\e^{\i\theta_2}\right)\right]\,, \\[.5em]
V_0&=-4\sin^2 ak\cdot\det S\,.
\end{align*}

\subsection{Linearly growing bands and gaps, or absence of gaps}

If we divide the above spectral condition by $k^2$, we can write
it in the asymptotic form $V_2(ak)=\O\left(k^{-1}\right)$,
explicitly
\begin{multline*}
-4\cos^2 ak\left[|t_{11}|^2+|t_{22}|^2+|t_{12}|^2+|t_{21}|^2\right]+4\sin^2 ak\left[1+\left|t_{11}t_{22}-t_{12}t_{21}\right|^2\right]\\
+8\cos ak\left[-\Re\left((t_{11}\overline{t_{21}}+t_{12}\overline{t_{22}})\e^{\i\theta_1}\right)+
\Re\left((t_{22}\overline{t_{21}}+\overline{t_{11}}t_{12})\e^{\i\theta_2}\right)\right]\\ +8\Re\left[t_{11}\overline{t_{22}}\e^{\i(\theta_1-\theta_2)}\right]
+8\Re\left[t_{12}\overline{t_{21}}\e^{\i(\theta_1+\theta_2)}\right]=\O\left(k^{-1}\right)\,,
\end{multline*}
from which it is possible to obtain the ``generic'' spectral
behaviour. Let us examine the \emph{lhs} of the last relation. To
this aim, we denote
\begin{gather*}
K_c:=4\left(|t_{11}|^2+|t_{22}|^2+|t_{12}|^2+|t_{21}|^2\right) \\
K_s:=4\left(1+\left|t_{11}t_{22}-t_{12}t_{21}\right|^2\right) \\
L_c(\theta_1,\theta_2):=8\left[-\Re\left((t_{11}\overline{t_{21}}+t_{12}\overline{t_{22}})\e^{\i\theta_1}\right)+
\Re\left((t_{22}\overline{t_{21}}+\overline{t_{11}}t_{12})\e^{\i\theta_2}\right)\right] \\
L(\theta_1,\theta_2):=8\Re\left(t_{11}\overline{t_{22}}\e^{\i(\theta_1-\theta_2)}\right)
+8\Re\left(t_{12}\overline{t_{21}}\e^{\i(\theta_1+\theta_2)}\right)
\end{gather*}
which allows us to write the coefficient $V_2\equiv
V_2(ak,\theta_1,\theta_2)$ as follows
$$
V_2(x,\theta_1,\theta_2)=-K_c\cos^2 x+K_s\sin^2 x+\cos x\cdot L_c(\theta_1,\theta_2)+L(\theta_1,\theta_2)\,.
$$
To examine the spectral asymptotics, the following functions,
\begin{gather*}
V_2^+(x):=\max\left\{\left.
V_2(x,\theta_1,\theta_2)\,\right|\,\theta_1,\theta_2\in(-\pi,\pi]\right\}\,, \\
V_2^-(x):=\min\left\{\left.
V_2(x,\theta_1,\theta_2)\,\right|\,\theta_1,\theta_2\in(-\pi,\pi]\right\}\,,
\end{gather*}
will be essential, since the spectral condition can be expressed,
up to an error of order of $\O\left(k^{-1}\right)$, by the
inequalities,
\begin{equation}\label{SPnerov}
V_2^+(ak)>0 \quad \wedge \quad V_2^-(ak)<0\,.
\end{equation}
As we will see below, the following two constants will play an important role:
\begin{gather*}
L_0^+:=8\max\left\{\Re\left((t_{11}\overline{t_{21}}+t_{12}\overline{t_{22}})\e^{\i\theta_1}\right)+
\Re\left((t_{22}\overline{t_{21}}+\overline{t_{11}}t_{12})\e^{\i\theta_2}\right)+\right.\\
\left.\left.+\Re\left(t_{11}\overline{t_{22}}\e^{\i(\theta_1-\theta_2)}\right)
+\Re\left(t_{12}\overline{t_{21}}\e^{\i(\theta_1+\theta_2)}\right)\,\right|\,\theta_1,\theta_2\in(-\pi,\pi]\right\}\,, \\
L_\frac{\pi}{2}^-:=8\min\left\{\left. \Re\left(t_{11}\overline{t_{22}}\e^{\i(\theta_1-\theta_2)}\right)
+\Re\left(t_{12}\overline{t_{21}}\e^{\i(\theta_1+\theta_2)}\right)\,\right|\,\theta_1,\theta_2\in(-\pi,\pi]\right\}\,;
\end{gather*}
it is easy to see that
$L_\frac{\pi}{2}^- =-8|t_{11}t_{22}| -8|t_{12}t_{21}|$.

With this preliminary we are going to formulate and prove a claim
which will be useful not only here, but also at other places
further on.

\begin{prop}\label{LinNez}
Let
$$
F(\theta_1,\theta_2)=\Re\left(A_1\e^{\i\theta_1}\right)+\Re\left(A_2\e^{\i\theta_2}\right)
+\Re\left(A_3\e^{\i(\theta_1-\theta_2)}\right)+\Re\left(A_4\e^{\i(\theta_1+\theta_2)}\right)\,,
$$
where the coefficients $A_j$, $j=1,2,3,4$, are independent of
$\theta_1,\theta_2$. Then the range of this expression,
$\mathcal{F}:=\left\{\left.F(\theta_1,\theta_2)\,\right|\,\theta_1,\theta_2\in(-\pi,\pi]\right\}$,
is an interval which is non-degenerate if and only if there is an
index $j\in\{1,2,3,4\}$ such that $A_j\neq0$.
\end{prop}
\pf Since $F$ is continuous and $(-\pi,\pi]^2$ is connected,
$\mathcal{F}$ is an interval. To finish the proof it remains to
check that a constant $C\in\R$ such that
\begin{equation}\label{FC}
F(\theta_1,\theta_2)=C \qquad \text{for all} \quad (\theta_1,\theta_2)\in(-\pi,\pi]^2
\end{equation}
exists if and only if $A_j=0$ for all $j=1,2,3,4$.

Consider a fixed $\theta\in\R$ and a number $A\in\C$ such that
$\arg A=\alpha$, then $\Re(A\e^{\i\theta})
=\Re(|A|\e^{\i\alpha}\e^{\i\theta})=|A|\cos(\alpha+\theta)$. We
apply this idea to rewrite~\eqref{FC} as
$$
|A_1|\cos(\alpha_1+\theta_1)+|A_2|\cos(\alpha_2+\theta_2)
+|A_3|\cos(\alpha_3+\theta_1+\theta_2+|A_4|\cos(\alpha_4+\theta_1-\theta_2)-C=0
$$
for all $(\theta_1,\theta_2)\in(-\pi,\pi]^2$. It is easy to see
that the $5$-tuple
$$
1,\cos(\alpha_1+\theta_1),\cos(\alpha_2+\theta_2),
\cos(\alpha_3+\theta_1+\theta_2),\cos(\alpha_4+\theta_1-\theta_2)
$$
is a linearly independent set of functions on $[0,2\pi)^2$,
therefore~\eqref{FC} is satisfied if and only if
$A_1=A_2=A_3=A_4=C=0$. \pfk

To solve the spectral conditions~\eqref{SPnerov}, we need to know
basic characteristics of the functions $V_2^+(x)$ and $V_2^-(x)$
involved in it.

\begin{prop}\label{m2.prop}
The functions $V_2^+(x)$ and $V_2^-(x)$ have the following properties:
\begin{itemize}
\item[\textit{(i)}] Both $V_2^+(x)$, $V_2^-(x)$ are $\pi$-periodic and satisfy
$$
V_2^+\left(\frac{\pi}{2}-x\right)=V_2^+(x)\,,\quad V_2^-\left(\frac{\pi}{2}-x\right)=V_2^-(x)\,.
$$
\item[\textit{(ii)}] Function $V_2^-(x)$ is increasing in
$\left[0,\frac{\pi}{2}\right]$ and
$$
V_2^-(0)<0\,,\qquad V_2^-\left(\frac{\pi}{2}\right)=K_s+L_\frac{\pi}{2}^-\,.
$$
\item[\textit{(iii)}] It holds $V_2^+(0)=-K_c+L_0^+$, and there is
a number $x_0\in\left[0,\frac{\pi}{2}\right]$ such that
\begin{itemize}
\item $V_2^+$ is increasing in $\left[0,x_0\right]$,
\item $V_2^+(x)>0$ for all $x\in\left[x_0,\frac{\pi}{2}\right]$\,.
\end{itemize}
\item[\textit{(iv)}] If at least two entries of $T$ are nonzero,
then $V_2^+(x)>V_2^-(x)$ for all $x\in\left(0,\frac{\pi}{2}\right)$.
\end{itemize}
\end{prop}
\emph{Proof} of this proposition is technical and slightly long, and can be found
in the appendix.

\medskip

Using the above result we can characterize the spectrum.

\begin{prop}\label{m2.p3}
If at least two entries of $T$ are nonzero, then
the following holds true:
\begin{itemize}
\item If $L_0^+>K_c$ and $L_\frac{\pi}{2}^-<-K_s$, there is a
$k_0>0$ such that the interval $[k_0^2,+\infty)$ is in the
spectrum.
\item If $L_0^+<K_c$, the spectrum has asymptotically gaps of the
form
$$
\left(\frac{n^2\pi^2}{a^2}-\frac{2bn\pi}{a^2}+\O(1),
\frac{n^2\pi^2}{a^2}+\frac{2bn\pi}{a^2}+\O(1)\right)\,,
$$
where $b\in\left(0,\frac{\pi}{2}\right)$ is the number uniquely
determined by the condition $V_2^+(b)=0$.
\item If
$L_\frac{\pi}{2}^->-K_s$, the spectrum has asymptotically gaps of the form
$$
\left(\left(n+\frac{1}{2}\right)^2\frac{\pi^2}{a^2}-\frac{2cn\pi}{a^2}+\O(1),
\left(n+\frac{1}{2}\right)^2\frac{\pi^2}{a^2}+\frac{2cn\pi}{a^2}+\O(1)\right)\,,
$$
where $c\in\left(0,\frac{\pi}{2}\right)$ is uniquely determined by
the condition $V_2^-\left(\frac{\pi}{2}-c\right)=0$.
\end{itemize}
\end{prop}

\pf The argument is based on Proposition~\ref{m2.prop}
\textit{(i)}-\textit{(iv)}. First we notice that if $L_0^+>K_c$,
then $V_2^+(x)>0$ for all $x\in\left[0,\frac{\pi}{2}\right]$
according to \textit{(ii)}, and thus for all $x\in[0,+\infty)$ due
to \textit{(i)}. Similarly, if $L_\frac{\pi}{2}^-<-K_s$, then
$V_2^-(x)<0$ for all $x\in\left[0,\frac{\pi}{2}\right]$ according
to \textit{(iii)}, and thus for all $x\in[0,+\infty)$ due to
\textit{(i)}. Consequently, for both $L_0^+>K_c$ and
$L_\frac{\pi}{2}^-<-K_s$, the asymptotic spectral
condition~\eqref{SPnerov} is satisfied for all $k$.

If $L_0^+<K_c$, then by virtue of \textit{(ii)} there is a
$b\in\left(0,\frac{\pi}{2}\right)$ such that $V_2^+(x)<0$ on
$[0,b)$ and $V_2^+(x)>0$ on $\left(b,\frac{\pi}{2}\right)$. Then
the first inequality of~\eqref{SPnerov} is not satisfied on
$\left(0,\frac{b}{a}\right)$, and we infer from \textit{(i)} that
it is not satisfied on each interval
$\left(\frac{n\pi}{a}-\frac{b}{a},\frac{n\pi}{a}+\frac{b}{a}\right)$.

In the same vein, the inequality $L_\frac{\pi}{2}^->-K_s$ in
combination with \textit{(iii)} implies existence of a
$c\in\left(0,\frac{\pi}{2}\right)$ such that $V_2^-(x)<0$ on
$[0,\frac{\pi}{2}-c)$ and $V_2^-(x)>0$ on
$\left(\frac{\pi}{2}-c,\frac{\pi}{2}\right)$. The second
inequality of~\eqref{SPnerov} is then not satisfied on
$\left(\frac{\pi}{2a}-\frac{c}{a},\frac{\pi}{2a}+\frac{c}{a}\right)$,
and by \textit{(i)} it is not satisfied on each interval
$\left(\left(n+\frac{1}{2}\right)\frac{\pi}{a}
-\frac{c}{a},\left(n+\frac{1}{2}\right)\frac{\pi}{a}+\frac{c}{a}\right)$.

Finally, it follows from \textit{(iv)} that if the matrix $T$ has at least two
non-vanishing entries, we have $b<\frac{\pi}{2}-c$ and the
gaps cannot overlap asymptotically. \pfk

\subsection{Further particular cases}

Proposition~\ref{m2.p3} applies if at least two entries of $T$ are
nonzero. If this is not the case, the spectral condition
simplifies significantly and one of the following situations
occurs:
\begin{itemize}
\item $s_{12}=0$. With such boundary conditions, the lattice is
decomposed into separated edges or pairs of edges, and
consequently the spectrum is pure point. As above in similar cases
it is infinitely degenerate, of course, but the contribution from
each edge pair has the usual semiclassical behaviour: the number
of eigenvalues not exceeding $k^2$ has Weyl asymptotics with the
leading term $\frac a\pi$.
\item $s_{12}\neq0$ and $T=0$. The spectrum contains points
$\frac{n^2\pi^2}{a^2}$ and bands of asymptotically constant width
in the vicinity of $\frac{n^2\pi^2}{a^2}$; the points
$\frac{n^2\pi^2}{a^2}$ may or may not lie in the bands. The gaps
grows linearly as $n\to\infty$.
\item $s_{12}\neq0$ and $T\neq0$. The spectrum asymptotically
consists of bands located in the vicinity of the points
$\left(\frac{n\pi}{a}+\frac{1}{2}\arccos\frac{1-|t|^2}{1+|t|^2}\right)^2$,
where $t$ is the nonzero entry of $T$. They are of asymptotically
constant width while the gaps are linearly growing as the band
index $n\to\infty$.
\end{itemize}
It remains to deal with the special cases $L_0^+=K_c$,
$L_\frac{\pi}{2}^-=-K_s$ left out in Proposition~\ref{m2.p3}.  We
will not go into much detail here, but it is worth to solve this
situation in the particular case of \emph{scale-invariant
coupling}, i.e. for $S=0$. It turns out that:
\begin{itemize}
\item If $S=0$ and $L_0^+=K_c$, then there are no gaps around
$\frac{n^2\pi^2}{a^2}$. \item If $S=0$ and
$L_\frac{\pi}{2}^-=-K_s$, then there are no gaps around
$\left(n+\frac{1}{2}\right)^2\frac{\pi^2}{a^2}$.
\end{itemize}



\section{The case of  $m=3$}

In this case we have a Hermitean $S$ and complex numbers $t_j$
determining
$$
B=\left(\begin{array}{cccc}
1 & 0 & 0 & t_1 \\
0 & 1 & 0 & t_2 \\
0 & 0 & 1 & t_3 \\
0 & 0 & 0 & 0
\end{array}\right)\,,\quad
A=
-\left(\begin{array}{cccc}
s_{11} & s_{12} & s_{13} & 0 \\
\overline{s_{12}} & s_{22} & s_{23} & 0 \\
\overline{s_{13}} & \overline{s_{23}} & s_{33} & 0 \\
-\overline{t_1} & -\overline{t_2} & -\overline{t_3} & 1
\end{array}\right)\,.
$$
In contrast to the previous case, there is no problem with the
renumbering invariance, because the edge numbering
can be changed if necessary by a trivial rotation of the lattice. A direct
calculation of the determinant in \eqref{ObecSp} yields the
spectral condition,
$$
V_3\cdot k^3+V_2\cdot k^2+V_1\cdot k+V_0=0\,,
$$
where $V_j$, $j=0,1,2,3$, are expressions depending on $ak$, on
the entries of the matrices $S$, $T$, and on the quasimomentum
components $\theta_1,\theta_2$, given explicitly by
\begin{align*}
V_3&=4\sin ak\left[\left(1+|t_1|^2+|t_2|^2+|t_3|^2\right)\cos ak
+2\Re(t_1\overline{t_2}\e^{\i\theta_1})-2\Re(t_3\e^{\i\theta_2})\right]\,,\\[.5em]
V_2&=
4\cos^2 ak\left[-(s_{11}+s_{22})\left(1+|t_3|^2\right)-s_{33}(|t_1|^2+|t_2|^2)
+2\Re\left((s_{13}\overline{t_1}+s_{23}\overline{t_2})t_3\right)\right]\\
&+4\sin^2 ak\left[s_{11}|t_2|^2+s_{22}|t_1|^2+s_{33}-2\Re\left(s_{12}\overline{t_1}t_2\right)\right]\\
&+8\cos ak\left[\Re\left((-s_{12}+\overline{s_{23}}t_1\overline{t_3}+s_{13}\overline{t_2}t_3-s_{12} |t_3|^2-s_{33}t_1\overline{t_2})\e^{\i\theta_1}\right)\right.\\
&\left.\;\qquad\qquad+\Re\left(\left((s_{11}+s_{22})t_3-\overline{s_{13}}t_1-\overline{s_{23}}t_2\right)
\e^{\i\theta_2}\right)\right]\\
&+8\Re\left((s_{12}t_3-\overline{s_{23}}t_1)\e^{\i(\theta_1+\theta_2)}\right)
+8\Re\left((s_{12}\overline{t_3}-s_{13}\overline{t_2})\e^{\i(\theta_1-\theta_2)}\right)\,,\\[.5em]
V_1&=4\sin ak\cos ak\left[(|s_{12}|^2-s_{11}s_{22})(1+|t_{3}|^2)+(|s_{13}|^2-s_{11}s_{33})(1+|t_{2}|^2)\right.\\
&\qquad\qquad\qquad\qquad+(|s_{23}|^2-s_{22}s_{33})(1+|t_{1}|^2)+\\
&\left.+2\Re\left((s_{12}s_{33}-s_{13}\overline{s_{23}})\overline{t_1}t_2\right)
+2\Re\left((s_{23}s_{11}-s_{13}\overline{s_{12}})\overline{t_2}t_3\right)
+2\Re\left((s_{13}s_{22}-s_{12}s_{23})\overline{t_1}t_3\right)\right]\\
&+8\sin ak\cdot\Re\left((s_{13}\overline{s_{23}}-s_{33}s_{12})\e^{\i\theta_1}\right)\\
&+8\sin ak\cdot\Re\left[\left((\overline{s_{12}}\overline{s_{23}}-s_{22}\overline{s_{13}})t_1+(s_{12}\overline{s_{13}}-s_{11}\overline{s_{23}})t_2+(s_{11}s_{22}-|s_{12}|^2)t_3\right)\e^{\i\theta_2}\right]\,,\\[.5em]
V_0&=-4\sin^2 ak\cdot\det S\,.
\end{align*}

The case $m=3$ is probably the most interesting one. While in
the one-dimensional case we have bands and gaps which are
asymptotically either of constant width or linearly growing with
the band index, a square lattice with $m=3$ exhibits a
considerably richer spectral behaviour. The band are here of two
types, mutually interlaced:
\begin{itemize}
\item ``even'' bands in the vicinity of $\frac{n^2\pi^2}{a^2}$
with $n\in\N$,
\item ``odd'' bands in the vicinity of
$\left(n+\frac{1}{2}\right)^2\frac{\pi^2}{a^2}$ with $n\in\N$,
\end{itemize}
In this section we focus on the generic situation,
$|t_1|\neq|t_2|$ or $|t_3|\neq1$, when we will be able to describe
several interesting asymptotic types. The other possibility,
$|t_1|=|t_2|\wedge |t_3|=1$, will be briefly commented at the end.

Let us start with the spectral condition in the form
$V_3=\O(k^{-1})$, i.e.
\begin{equation}\label{m3.1}
4\sin ak\left[\left(1+|t_1|^2+|t_2|^2+|t_3|^2\right)\cos ak
+2\Re(t_1\overline{t_2}\e^{\i\theta_1})
-2\Re(t_3\e^{\i\theta_2})\right]=\O(k^{-1})\,.
\end{equation}
The \emph{lhs} of the last equation should be close to zero for
large $k$, which can be achieved either for small absolute value
of $\sin ak$ or for small absolute value of the expression in the
brackets. These possibilities refer to the ``even'' and ``odd''
bands mentioned above, respectively; they will be described in
detail below. Unless stated otherwise, we suppose that
$|t_1|\neq|t_2|\vee |t_3|\neq1$.


\subsection{Generic case: asymptotically constant spectral bands around $\frac{n^2\pi^2}{a^2}$}\label{m3.s2}

We start with the bands corresponding to small values of $|\sin
ak|$. Let us consider the spectral condition in the form
$V_3+\frac{V_2}{k}=\O(k^{-2})$, divide it by four and rewrite it
as
\begin{multline}\label{m3.4}
\sin ak\left[\left(1+|t_1|^2+|t_2|^2+|t_3|^2\right)\cos ak
+2\Re(t_1\overline{t_2}\e^{\i\theta_1})-2\Re(t_3\e^{\i\theta_2})\right.\\
\left.-\frac{\sin ak}{k}\left(s_{11}|t_2|^2+s_{22}|t_1|^2+s_{33}-2\Re\left(s_{12}\overline{t_1}t_2\right)\right)\right]\\
=-\frac{\cos^2 ak}{k}\left[-(s_{11}+s_{22})\left(1+|t_3|^2\right)-s_{33}(|t_1|^2+|t_2|^2)
+2\Re\left((s_{13}\overline{t_1}+s_{23}\overline{t_2})t_3\right)\right]\\
-2\frac{\cos ak}{k}\left[\Re\left((-s_{12}+\overline{s_{23}}t_1\overline{t_3}+s_{13}\overline{t_2}t_3-s_{12} |t_3|^2-s_{23}t_1\overline{t_2})\e^{\i\theta_1}\right)\right.\\
\left.+\Re\left(\left((s_{11}+s_{22})t_3-\overline{s_{13}}t_1-\overline{s_{23}}t_2\right)
\e^{\i\theta_2}\right)\right]\\
-\frac{2}{k}\Re\left((s_{12}t_3-\overline{s_{23}}t_1)\e^{\i(\theta_1+\theta_2)}\right)
-\frac{2}{k}\Re\left((s_{12}\overline{t_3}-s_{13}\overline{t_2})\e^{\i(\theta_1-\theta_2)}\right)+\O(k^{-2})\,.
\end{multline}
We restrict our considerations to the values of $k$ such that
$|\cos ak|>c$ where $c$ is a constant satisfying
$\frac{2(|t_1t_2|+|t_3|)}{1+|t_1|^2+|t_2|^2+|t_3|^2}<c<1$; recall
that such $c$ exists in view of the initial assumption
$|t_1|\neq|t_2|\vee|t_3|\neq1$. Due to this restriction, the absolute value of the term
in the brackets at the \emph{lhs} of~\eqref{m3.4} can be asymptotically estimated
from below by a positive constant, whence it is easy to
see that high in the spectrum we have $\sin ak=\O(k^{-1})$. We set
$$
k=\frac{n\pi}{a}+\frac{d}{n}\,,
$$
then $\sin ak=(-1)^n\cdot\frac{ad}{n} +\O(n^{-3})$, $\cos
ak=(-1)^n+\O(n^{-2})$ and $\frac{1}{k}=\frac{a}{n\pi}+\O(n^{-2})$.
We substitute from these relations into \eqref{m3.4}, divide both
sides by the expression in the brackets at the \emph{lhs} and
after elementary manipulations we arrive at
\begin{multline*}
\frac{ad}{n}\cdot\left[\left(1+|t_1|^2+|t_2|^2+|t_3|^2\right)+2(-1)^n\Re(t_1\overline{t_2}\e^{\i\theta_1})-2(-1)^n\Re(t_3\e^{\i\theta_2})\right]=\\
=-\frac{a}{n\pi}\left[-(s_{11}+s_{22})\left(1+|t_3|^2\right)-s_{33}(|t_1|^2+|t_2|^2)
+2\Re\left((s_{13}\overline{t_1}+s_{23}\overline{t_2})t_3\right)\right]\\
-2\frac{(-1)^na}{n\pi}\left[\Re\left((-s_{12}+\overline{s_{23}}t_1\overline{t_3}+s_{13}\overline{t_2}t_3-s_{12} |t_3|^2-s_{23}t_1\overline{t_2})\e^{\i\theta_1}\right)\right.\\
\left.+\Re\left(\left((s_{11}+s_{22})t_3-\overline{s_{13}}t_1-\overline{s_{23}}t_2\right)
\e^{\i\theta_2}\right)\right]\\
-\frac{2a}{n\pi}\Re\left((s_{12}t_3-\overline{s_{23}}t_1)\e^{\i(\theta_1+\theta_2)}\right)
-\frac{2a}{n\pi}\Re\left((s_{12}\overline{t_3}-s_{13}\overline{t_2})\e^{\i(\theta_1-\theta_2)}\right)+\O(n^{-2})\,.
\end{multline*}
We observe that all the terms $(-1)^n$ can be ``absorbed'' into
$\theta_1$ and $\theta_2$ by the shift
$(\theta_1,\theta_2)\mapsto(\theta_1+n\pi,\theta_2+n\pi)$, and thus
neglected. Therefore the asymptotic condition can be written in
the form
$$
d=\frac{1}{\pi}\cdot\frac{-W_2(S,T,\theta_1,\theta_2)}{W_3(T,\theta_1,\theta_2)}+\O(n^{-1})
$$
for
$$
W_3(T,\theta_1,\theta_2):=\left(1+|t_1|^2+|t_2|^2+|t_3|^2\right)
+2(-1)^n\Re(t_1\overline{t_2}\e^{\i\theta_1})-2(-1)^n\Re(t_3\e^{\i\theta_2})\,,
$$
\begin{multline*}
W_2(S,T,\theta_1,\theta_2):=-(s_{11}+s_{22})\left(1+|t_3|^2\right)-s_{33}(|t_1|^2+|t_2|^2)
+2\Re\left((s_{13}\overline{t_1}+s_{23}\overline{t_2})t_3\right)\\
+2\Re\left((-s_{12}+\overline{s_{23}}t_1\overline{t_3}+s_{13}\overline{t_2}t_3-s_{12} |t_3|^2-s_{33}t_1\overline{t_2})\e^{\i\theta_1}\right)\\
+2\Re\left(\left((s_{11}+s_{22})t_3-\overline{s_{13}}t_1-\overline{s_{23}}t_2\right)\e^{\i\theta_2}\right)\\
+2\Re\left((s_{12}t_3-\overline{s_{23}}t_1)\e^{\i(\theta_1+\theta_2)}\right)
+2\Re\left((s_{12}\overline{t_3}-s_{13}\overline{t_2})\e^{\i(\theta_1-\theta_2)}\right)\,.
\end{multline*}
Now we denote
$$
\mathcal{D}:=\left\{\left.\frac{-W_2(S,T,\theta_1,\theta_2)}{W_3(T,\theta_1,\theta_2)}
\,\right|\,\theta_1,\theta_2\in(-\pi,\pi]\right\}\,;
$$
note that $\mathcal{D}$ is an interval which is bounded due to the
condition $|t_1|\neq|t_2|\vee|t_3|\neq1$. The spectral bands are
then described in terms of $\mathcal{D}$ as follows,
\begin{equation} \label{spec7.1}
k=\frac{n\pi}{a}+\frac{d}{n\pi}+\O(n^{-2})\,,\quad
d\in\mathcal{D}\,,
\end{equation}
which yields the allowed energy values
$$
k^2=\frac{n^2\pi^2}{a^2}+2\,\frac{d}{a}+\O(n^{-1})\,,\quad
d\in\mathcal{D}\,.
$$
Consequently, these bands are of asymptotically constant width as
$n\to\infty$.

Furthermore, it is easy to show that if
\begin{itemize}
\item[(a)] $T=0$ and $s_{12}=0$, or
\item[(b)] $t_1t_2=0$, $t_3=0$ and $S$ is diagonal,
\end{itemize}
then the fraction
$\frac{-W_2(S,T,\theta_1,\theta_2)}{W_3(T,\theta_1,\theta_2)}$ is
independent of $(\theta_1,\theta_2)$, and thus the interval
$\mathcal{D}$ shrinks to a point. These cases need a special approach.

One can check that (b) refers to the situation when the lattice is
decomposed either to individual edges (if $T=0$) or to pairs of
edges (if $T\neq0$), with infinitely degenerate eigenvalues in the
vicinity of $\left(\frac{n\pi}{a}\right)^2$. The case (a) is much
more interesting and we will analyze it below.


\subsection{Spectral bands around $\frac{n^2\pi^2}{a^2}$ shrinking as $n^{-2}$}

Under the assumptions of the case (a) above the spectral
condition simplifies to
\begin{multline*}
4k^3\sin ak\cos ak+4k^2\left[-(s_{11}+s_{22})\cos^2 ak+s_{33}\sin^2 ak\right]\\
+4k\sin ak\left[(|s_{13}|^2+|s_{23}|^2-s_{11}s_{22}-(s_{11}+s_{22})s_{33})\cos ak
+2\Re\left(s_{13}\overline{s_{23}}\e^{\i\theta_1}\right)\right]\\
-4\sin^2 ak\cdot\det S=0\,.
\end{multline*}
The spectrum is given by \eqref{spec7.1} with
$\mathcal{D}=\left\{\frac{a}{\pi}(s_{11}+s_{22})\right\}$. Since
the latter is a one-point set, the spectral band is determined by
the next term in the expansion. We set
$$
k=\frac{n\pi}{a}+\frac{a}{n\pi}(s_{11}+s_{22})+\frac{d'}{n^2}\,,
$$
and performing a calculation similar to that of Sec.~\ref{m3.s2},
we arrive at the expression for $d'$. Surprisingly, $d'$ itself
behaves like $n^{-1}$, i.e. the last term above is of order of
$\O(n^{-3})$,
$$
\frac{d'}{n^2}=\frac{a(s_{11}+s_{22})}{n^3\pi^3}\left[-(s_{11}+s_{22})
+a\left(\frac{s_{11}^2+s_{22}^2}{2}+2s_{11}s_{22}+|s_{13}|^2+|s_{23}|^2\right)
-2a\Re\left(s_{13}\overline{s_{23}}\e^{\i\theta_1}\right)\right]\,.
$$
Thus $k^2$  belongs asymptotically to the spectrum if
\begin{multline*}
k=\frac{n\pi}{a}+\frac{s_{11}+s_{22}}{n\pi}+\frac{a(s_{11}+s_{22})}{n^3\pi^3}\left[-(s_{11}+s_{22})
+a\left(\frac{s_{11}^2+s_{22}^2}{2}+2s_{11}s_{22}+|s_{13}|^2+|s_{23}|^2\right)\right]\\
+\frac{2a^2}{n^3\pi^3}|s_{11}+s_{22}|\cdot|s_{13}s_{23}|\cos\vartheta+\O(n^{-4})\qquad
\text{for $\vartheta\in[0,2\pi)$}\,.
\end{multline*}
The left and right endpoint of the spectral band are then given as
$k_L^2$ and $k_R^2$ with $\vartheta_L=\pi$ and $\vartheta_R=0$,
respectively, hence the band width behaves like
$$
k_R^2-k_L^2=(k_R-k_L)(k_R+k_L)
=8\frac{a}{n^2\pi^2}|s_{11}+s_{22}|\cdot|s_{13}s_{23}| +
\O(n^{-3})
$$
in the asymptotic regime, $n\to\infty$.

Note that there are two exceptional cases here, namely:
\begin{itemize}
\item $s_{13}=0$ or $s_{23}=0$. The spectral condition is
independent of $\theta_1,\theta_2$, thus it has only point
solutions.
\item $s_{11}+s_{22}=0$. The spectral condition factorizes, $\sin
ak\left[\cos ak+\O\left(\frac{1}{k}\right)\right]=0$, producing no
(true) band in the vicinity of $\frac{n^2\pi^2}{a^2}$.
\end{itemize}


\subsection{Generic case: linearly growing spectral bands around $\left(n+\frac{1}{2}\right)^2\frac{\pi^2}{a^2}$}\label{m3.s}

Now we proceed to the ``odd'' bands, i.e. those corresponding to
asymptotically small values of the expression
$\left|\left(1+|t_1|^2+|t_2|^2+|t_3|^2\right)\cos
ak+2\Re(t_1\overline{t_2}\e^{\i\theta_1})-2\Re(t_3\e^{\i\theta_2})\right|$.
Their structure can be derived directly from the spectral
condition~\eqref{m3.1}, we just replace for the sake of simplicity
the term
$2\Re(t_1\overline{t_2}\e^{\i\theta_1})-2\Re(t_3\e^{\i\theta_2})$
by $2(|t_1t_2|+|t_3|)\cos\vartheta$ in analogy with what we did in
Section~\ref{m1.LinGr}, cf. Observation~\ref{m1.o1}:
$$
4\sin ak\left[\left(1+|t_1|^2+|t_2|^2+|t_3|^2\right)\cos ak
+2(|t_1t_2|+|t_3|)\cos\vartheta\right]=\O(k^{-1})
$$
Suppose that $|t_1t_2|+|t_3|\neq0$ (the case $t_1t_2=t_3=0$ is
postponed to Sec.~\ref{m3.ss} below), then we may divide the
equation by $8(|t_1\overline{t_2}|+|t_3|)$ obtaining
\begin{equation}\label{m3.2}
\sin
ak\left[\frac{1+|t_1|^2+|t_2|^2+|t_3|^2}{2(|t_1t_2|+|t_3|)}\cos ak
+\cos\vartheta\right]=\O(k^{-1})\,.
\end{equation}
Since the situation when $|\sin ak|$ is very small has been
treated in the part devoted to ``even'' bands, here we adopt the
premise that $|\sin ak|$ is large enough. More precisely, we
restrict our considerations to such values $k$ that $|\sin ak|>c$,
where $c$ is a constant satisfying $0<c<1$, the exact value plays
no role. Then \eqref{m3.2} can be rewritten as
$$
\cos\vartheta=\frac{1+|t_1|^2+|t_2|^2+|t_3|^2}{2(|t_1t_2|+|t_3|)}\cos
ak+\O(k^{-1})\,.
$$
Such a condition has a solution for some parameter $\vartheta$
if the
modulus of the rhs \emph{rhs} is smaller than one, i.e. for
\begin{equation}\label{m3.3}
|\cos
ak|<\frac{2(|t_1t_2|+|t_3|)}{1+|t_1|^2+|t_2|^2+|t_3|^2}+\O(k^{-1})\,.
\end{equation}
In analogy with Sec.~\ref{m1.LinGr}, we infer that the spectral
bands are neigbourhoods of the points
$\left[\left(n+\frac{1}{2}\right)\frac{\pi}{a}\right]^2$. We set
$k=\left(n+\frac{1}{2}\right)\frac{\pi}{a}+d$, thus $|\cos
ak|=|\sin ad|$ and $\frac{1}{k}$ of the $n$-th band solution is
$\O(n^{-1})$. Then \eqref{m3.3} can be rewritten as
$$
|\sin ad|<\frac{2(|t_1t_2|+|t_3|)}{1+|t_1|^2+|t_2|^2+|t_3|^2}+\O(n^{-1})\,.
$$
Recall that it follows from the initial assumption
$|t_1|\neq|t_2|\vee|t_3|\neq1$ that the first term at the
\emph{rhs} is smaller than one, cf. Observation~\ref{m1.o1}, hence
it makes sense to set $\Delta:=
\arcsin\frac{2(|t_1t_2|+|t_3|)} {1+|t_1|^2+|t_2|^2+|t_3|^2}\in\left(0,\frac{\pi}{2}\right)$. The
spectral bands are then given by
$$
\left(\left(n+\frac{1}{2}\right)^2\frac{\pi^2}{a^2}
-2\,\frac{n\pi}{a^2}\Delta+\O(1),\left(n+\frac{1}{2}\right)^2
\frac{\pi^2}{a^2}+2\,\frac{n\pi}{a^2}\Delta+\O(1)\right)\,,
$$
in the high-energy asymptotics, $n\to\infty$.


\subsection{Asymptotically constant spectral bands at $\left(n+\frac{1}{2}\right)^2\frac{\pi^2}{a^2}$}\label{m3.ss}

Let the matrix $T$ satisfy $t_1t_2=t_3=0$. We may suppose without
loss of generality that $t_2=0$. The substitution $t_3=t_2=0$ into
the spectral condition $V_3+\frac{V_2}{k}=\O(k^{-2})$ gives
\begin{multline}\label{m3.SP0}
\sin ak\cos ak\left(1+|t_1|^2\right)=\frac{\cos^2 ak}{k}\left(s_{11}+s_{22}+s_{33}|t_1|^2\right)
-\frac{\sin^2 ak}{k}\left(s_{22}|t_1|^2+s_{33}\right)\\
+\frac{2\cos
ak}{k}\left[\Re\left(s_{12}\e^{\i\theta_1}\right)+\Re\left(\overline{s_{13}}t_1\e^{\i\theta_2}\right)\right]
+\frac{2}{k}\Re\left(t_1\overline{s_{23}}\e^{\i(\theta_1+\theta_2)}\right)+\O(k^{-2})\,.
\end{multline}
We proceed in a way analogous to Sec.~\ref{m3.s2} setting
$k=\left(n+\frac{1}{2}\right)\frac{\pi}{a}+\delta$ and
substituting the expansions
$$
\cos ak=-(-1)^n\cdot a\delta+\O\left(\delta^3\right)\,,\qquad \sin
ak=(-1)^n+\O\left(\delta^2\right)\,,\qquad
\frac{1}{k}=\frac{a}{n\pi}+\O(n^{-2})
$$
into \eqref{m3.SP0}, which yields after a simple manipulation an
expression for $\delta$,
$$
\delta=\frac{1}{n\pi}\cdot\left[-\frac{s_{22}|t_1|^2+s_{33}}{1+|t_1|^2}
+\frac{2}{1+|t_1|^2}|\Re\left(t_1\overline{s_{23}}\e^{\i(\theta_1+\theta_2)}\right)\right]
+\O(n^{-2})\,.
$$
Hence the spectral bands are determined by the values of $k$
satisfying
$$
k=\left(n+\frac{1}{2}\right)\frac{\pi}{a}-\frac{1}{n\pi}\cdot\frac{s_{22}|t_1|^2+s_{33}}{1+|t_1|^2}
+\frac{1}{n\pi}\cdot\frac{2|t_1s_{23}|}{1+|t_1|^2}\cos\vartheta
+\O(n^{-2})
$$
so that
$$
k^2=\left(n+\frac{1}{2}\right)^2\frac{\pi^2}{a^2}+\frac{2}{a}\cdot\frac{s_{22}|t_1|^2+s_{33}}{1+|t_1|^2}
+\frac{2}{a}\cdot\frac{2|t_1s_{23}|}{1+|t_1|^2}\cos\vartheta
+\O(n^{-1})\,,\quad \vartheta\in[0,2\pi)\,.
$$
giving rise to asymptotically constant-width bands.

Note that the leading ``band-producing'' term may collapse to a
single point which happens if $t_1=0$ or $s_{23}=0$. The first
named situation will be discussed in detail in Sections
\ref{m3.subs.odd1}--\ref{m3.subs.odd3}, the second one will be
omitted, since the analysis is in principle similar.


\subsection{Spectral bands at $\left(n+\frac{1}{2}\right)^2\frac{\pi^2}{a^2}$ shrinking as $n^{-1}$}\label{m3.subs.odd1}

The bands in the vicinity of $\left(n+\frac{1}{2}\right)^2
\frac{\pi^2}{a^2}$ may shrink as $n^{-1}$ provided the matrices
$S$ and $T$ satisfy certain conditions in addition to $t_2=t_3=0$, namely $t_1=0$ and $s_{13}s_{23}\neq0$. We will demonstrate this fact in the first case, $t_1=0$.

We proceed as in Sec.~\ref{m3.ss}, but we take one more term
in the expansion of $k$, i.e. we set
$$
k=\left(n+\frac{1}{2}\right)\frac{\pi}{a}+\frac{s_{33}}{n\pi}+\delta\,,
$$
and substitute into the spectral condition in the form
$V_3+\frac{V_2}{k}+\frac{V_1}{k^2}=\O(k^{-3})$. A calculation
leads to the expression
$\delta=\frac{1}{n^2}\left[-\frac{s_{33}}{2\pi}
+\frac{2a}{\pi^2}\Re\left((s_{13}\overline{s_{23}})\e^{\i\theta_1}\right)\right]$
in the leading order, hence
$$
k=\left(n+\frac{1}{2}\right)\frac{\pi}{a}
+\frac{s_{33}}{n\pi}-\frac{s_{33}}{2n^2\pi}
+\frac{2a}{n^2\pi^2}|s_{13}s_{23}|\cos\vartheta+\O(n^{-3})\,,\quad
\vartheta\in[0,2\pi)\,.
$$
The band edges correspond to $\vartheta_L=\pi,0$, respectively,
and the band width equals
$$
\frac{8}{n\pi}|s_{13}s_{23}|+\O(n^{-2})\,,
$$
i.e. the bands shrink asymptotically as $n^{-1}$ unless $s_{13}=0$
or $s_{23}=0$ -- these cases will be discussed in the following sections.


\subsection{Spectral bands at $\left(n+\frac{1}{2}\right)^2\frac{\pi^2}{a^2}$ shrinking as $n^{-2}$}

If $T=0$ and $s_{23}=0$, the following spectral band asymptotics
can be derived,
$$
k=\left(n+\frac{1}{2}\right)\frac{\pi}{a}
+\frac{s_{33}}{n\pi}-\frac{s_{33}}{2n^2\pi}+
\frac{\mathcal{C}}{n^3}
-\frac{a}{n^3}\cdot\frac{s_{33}}{\pi^2}|s_{12}|\cos\vartheta+\O(n^{-4})\,,\quad
\vartheta\in[0,2\pi)\,,
$$
where $\mathcal{C}$ depends only on $S$ and $a$.
This determines bands of the widths
$$
\frac{4}{n^2\pi}|s_{33}s_{12}|+\O(n^{-3})\,,
$$
i.e. they shrink asymptotically as $n^{-2}$ providing that
$s_{33}\neq0$ and $s_{12}\neq0$.

\begin{pozn}
If $T=0$, $s_{23}=0$ and $s_{12}=0$, the spectral condition is independent of $\theta_1,\theta_2$, thus the spectrum is pure point.
\end{pozn}


\subsection{Spectral bands at $\left(n+\frac{1}{2}\right)^2\frac{\pi^2}{a^2}$ shrinking as $n^{-3}$}\label{m3.subs.odd3}

Let $T=0$ and $s_{23}=s_{33}=0$. Then the spectral bands are
characterized by the following asymptotical condition,
$$
k=\left(n+\frac{1}{2}\right)\frac{\pi}{a}
-\frac{a^2}{n^3}\cdot\frac{|s_{13}|^2s_{22}}{\pi^3}
+\frac{a^2}{n^4}\cdot\frac{3|s_{13}|^2s_{22}}{2\pi^3}
+\frac{a^3}{n^4}\cdot\frac{2|s_{13}|^2s_{22}}{2\pi^4}|s_{12}|\cos\vartheta+\O(n^{-5})
$$
with $\vartheta\in[0,2\pi)$, thus the band widths are given by
$$
8\,\frac{a^2}{n^3}\cdot\frac{|s_{13}|^2s_{22}}{2\pi^3}|s_{12}|\cos\vartheta+\O(n^{-4})\,,
$$
i.e. they shrink asymptotically as $n^{-3}$ providing
$s_{12}\neq0$, $s_{13}\neq0$ and $s_{22}\neq0$.

\begin{pozn}
Let $T=0$, $s_{23}=0$ and $s_{33}=0$. Then:
\begin{itemize}
\item If $s_{12}=0$, the spectral condition is independent of the
quasimomentum components $\theta_1,\theta_2$ so the spectrum is
pure point.
\item If $s_{13}=0$ or $s_{22}=0$, the spectral condition can be
factorized in the form $\cos ak\cdot\left[\sin
ak+\O\left(\frac{1}{n}\right)\right]$, and consequently, the band
in the vicinity of $\left(n+\frac{1}{2}\right)^2\frac{\pi^2}{a^2}$
collapses into a single point.
\end{itemize}
\end{pozn}


\subsection{The particular case $|t_1|=|t_2|\wedge |t_3|=1$}

This situation will not be treated in detail, but we will comment
on the prominent case $S=0$ which corresponds to the
\emph{scale-invariant vertex coupling}. If $S$ vanishes, the
spectral condition simplifies to $V_3=0$ which can be
rewritten as
$$
4\sin ak\left[\left(2+2|t_1|^2\right)\cos ak-2|t_1|^2\cos\vartheta_1-2\cos\vartheta_2\right]=0\,,
$$
where $\vartheta_1$
and $\vartheta_2$ are properly shifted $\theta_1,\theta_2$. 
The choice $\vartheta_1=\vartheta_2=ak$ sets the expression in the
brackets to zero, therefore the spectrum contains the positive
halfline.


\section{The case of  $m=4$}

Finally, we pass to the case which is generic from the viewpoint
of the boundary condition \eqref{bc:KS} when the matrix $B$ is
regular. Following the discussion in the opening we set
$$
B=\left(\begin{array}{cccc}
1 & 0 & 0 & 0 \\
0 & 1 & 0 & 0 \\
0 & 0 & 1 & 0 \\
0 & 0 & 0 & 1
\end{array}\right)\,,\quad
A=
-\left(\begin{array}{cccc}
s_{11} & s_{12} & s_{13} & s_{14} \\
\overline{s_{12}} & s_{22} & s_{23} & s_{24} \\
\overline{s_{13}} & \overline{s_{23}} & s_{33} & s_{34} \\
\overline{s_{14}} & \overline{s_{24}} & \overline{s_{34}} & s_{44}
\end{array}\right)
$$
There is obviously no problem with the renumbering the lattice edges.
A direct calculation of the determinant in \eqref{ObecSp} leads to
the spectral condition
$$
V_4\cdot k^4+V_3\cdot k^3+V_2\cdot k^2+V_1\cdot k+V_0=0\,,
$$
where $V_j$, $j=0,1,2,3,4$, are expressions depending on $ak$, on
entries of the matrix $S$, and on the quasimomentum components
$\theta_1,\theta_2$, given be the formul{\ae}
\begin{align*}
V_4&=-4\sin^2 ak\,,\\[.5em]
V_3&=4\sin ak\left[\left(s_{11}+s_{22}+s_{33}+s_{44}\right)\cos ak
+2\Re\left(s_{12}\e^{\i\theta_1}\right)+2\Re\left(s_{34}\e^{\i\theta_2}\right)\right]\,,\\[.5em]
V_2&=4\cos^2 ak\left(|s_{13}|^2-s_{11}s_{33}+|s_{14}|^2-s_{11}s_{44}+|s_{23}|^2-s_{22}s_{33}+|s_{24}|^2-s_{22}s_{44}\right)+\\
&+4\sin^2 ak\left(s_{11}s_{22}-|s_{12}|^2+s_{33}s_{44}-|s_{34}|^2\right)\\
&+8\cos ak\left[-(s_{33}+s_{44})\cdot\Re\left(s_{12}\e^{\i\theta_1}\right)+
\Re\left((s_{13}\overline{s_{23}}+s_{14}\overline{s_{24}})\e^{\i\theta_1}\right)\right.\\
&\left.-(s_{11}+s_{22})\cdot\Re\left(s_{34}\e^{\i\theta_2}\right)+
\Re\left((\overline{s_{13}}s_{14}+\overline{s_{23}}s_{24})\e^{\i\theta_2}\right)\right]\\
&+8\Re\left[(s_{14}\overline{s_{23}}-s_{12}s_{34})\e^{\i(\theta_1+\theta_2)}\right]
+8\Re\left[(s_{13}\overline{s_{24}}-s_{12}\overline{s_{34}})\e^{\i(\theta_1-\theta_2)}\right]\,,\\[.5em]
V_0&=-4\sin^2 ak\cdot\det S\,.
\end{align*}
The remaining term $V_1$ will not be
needed -- it is sufficient to know that it is bounded with respect
to the parameters. The spectral condition can be written as
\begin{equation}\label{m4.1}
\begin{split}
\sin^2 ak=&\frac{\sin ak}{k}\cdot\left[\left(s_{11}+s_{22}+s_{33}+s_{44}\right)\cos ak
+2\Re\left(s_{12}\e^{\i\theta_1}\right)+2\Re\left(s_{34}\e^{\i\theta_2}\right)\right]\\
&+\frac{\cos^2 ak}{k^2}\left(|s_{13}|^2-s_{11}s_{33}+|s_{14}|^2-s_{11}s_{44}+|s_{23}|^2-s_{22}s_{33}+|s_{24}|^2-s_{22}s_{44}\right)\\
&+\frac{2\cos ak}{k^2}\left[-(s_{33}+s_{44})\cdot\Re\left(s_{12}\e^{\i\theta_1}\right)+
\Re\left((s_{13}\overline{s_{23}}+s_{14}\overline{s_{24}})\e^{\i\theta_1}\right)\right.\\
&\left.-(s_{11}+s_{22})\cdot\Re\left(s_{34}\e^{\i\theta_2}\right)+
\Re\left((\overline{s_{13}}s_{14}+\overline{s_{23}}s_{24})\e^{\i\theta_2}\right)\right]\\
&+\frac{2}{k^2}\Re\left[\left((s_{14}\overline{s_{23}}-s_{12}s_{34})\e^{\i(\theta_1+\theta_2)}\right)
+8\Re\left((s_{13}\overline{s_{24}}-s_{12}\overline{s_{34}})\e^{\i(\theta_1-\theta_2)}\right)\right]\\
&+\frac{\sin^2 ak}{k^2}\left(s_{11}s_{22}-|s_{12}|^2+s_{33}s_{44}-|s_{34}|^2\right)
\\
&+\O\left(k^{-3}\right)\,.
\end{split}
\end{equation}
Since the \emph{rhs} is of the order of $\O(k^{-2})$, it follows
that the values of $k$ that solve the spectral condition are
asymptotically close to the points $\frac{n\pi}{a}$, $n\in\N$.
With this fact in mind it is convenient to express $k$ introducing
$d$ such that
\begin{equation}\label{m4.2}
k=\frac{n\pi}{a}+\frac{d}{n\pi}\,.
\end{equation}
Substituting this into \eqref{m4.1}, we get
after a simple manipulation,
\begin{multline}\label{m4.3}
d^2-d\cdot\left[s_{11}+s_{22}+s_{33}+s_{44}+2\cdot(-1)^n\cdot\Re\left(s_{12}\e^{\i\theta_1}\right)
+2\cdot(-1)^n\cdot\Re\left(s_{34}\e^{\i\theta_2}\right)\right]\\
-\left(|s_{13}|^2-s_{11}s_{33}+|s_{14}|^2-s_{11}s_{44}+|s_{23}|^2-s_{22}s_{33}+|s_{24}|^2-s_{22}s_{44}\right)\\
-2\cdot(-1)^n\cdot\left[-(s_{33}+s_{44})\cdot\Re\left(s_{12}\e^{\i\theta_1}\right)+
\Re\left((s_{13}\overline{s_{23}}+s_{14}\overline{s_{24}})\e^{\i\theta_1}\right)\right.\\
\left.-(s_{11}+s_{22})\cdot\Re\left(s_{34}\e^{\i\theta_2}\right)+
\Re\left((\overline{s_{13}}s_{14}+\overline{s_{23}}s_{24})\e^{\i\theta_2}\right)\right]\\
-2\Re\left[(s_{14}\overline{s_{23}}-s_{12}s_{34})\e^{\i(\theta_1+\theta_2)}\right]
-2\Re\left[(s_{13}\overline{s_{24}}-s_{12}\overline{s_{34}})\e^{\i(\theta_1-\theta_2)}\right]
=\O\left(n^{-1}\right)\,.
\end{multline}
We observe that the last relation has the structure
\begin{equation}\label{m4.4}
d^2+2p\cdot d+q=\O\left(n^{-1}\right)
\end{equation}
with the quantities $p,q$ depending only on $S$,
$\theta_1,\theta_2$ and the sign of $n$. A value $k^2>0$ with $k$
of the form~\eqref{m4.2} belongs to the spectrum if there are
$\theta_1,\theta_2$ such that the above condition is satisfied.
The solutions of~\eqref{m4.4} are given by
\begin{equation}\label{d12}
d_{1,2}=-p\pm\sqrt{p^2-q}+\O\left(n^{-1}\right)\,.
\end{equation}

\begin{prop}\label{m4.p1}
It holds $\max\{p^2-q\,|\,\theta_1,\theta_2\in(-\pi,\pi]\}\geq0$.
\end{prop}
\pf
Taking the value of $q$ from \eqref{m4.3} and $p^2$ arranged into the form
\begin{multline*}
p^2=\frac{1}{4}\left(s_{11}+s_{22}-s_{33}-s_{44}\right)^2+\left(s_{11}s_{33}+s_{11}s_{44}+s_{22}s_{33}+s_{22}s_{44}\right)+\\
(-1)^n\cdot\left(s_{11}+s_{22}+s_{33}+s_{44}\right)
\left[\Re\left(s_{12}\e^{\i\theta_1}\right)
+\Re\left(s_{34}\e^{\i\theta_2}\right)\right]+
\left[\Re\left(s_{12}\e^{\i\theta_1}\right)+\Re\left(s_{34}\e^{\i\theta_2}\right)\right]^2\,,
\end{multline*}
we find
\begin{multline}\label{p2q}
p^2-q=\frac{1}{4}\left(s_{11}+s_{22}-s_{33}-s_{44}\right)^2+\left[\Re\left(s_{12}\e^{\i\theta_1}\right)-\Re\left(s_{34}\e^{\i\theta_2}\right)\right]^2\\
+(-1)^n\cdot\Re\left[\left(\left(s_{11}+s_{22}-s_{33}-s_{44}\right)s_{12}+2\left(s_{13}\overline{s_{23}}
+s_{14}\overline{s_{24}}\right)\right)\e^{\i\theta_1}\right]\\
+(-1)^n\cdot\Re\left[\left(\left(-s_{11}-s_{22}+s_{33}+s_{44}\right)s_{34}+2\left(\overline{s_{13}}s_{14}
+\overline{s_{23}}s_{24}\right)\right)\e^{\i\theta_2}\right]\\
+|s_{14}|^2+|s_{23}|^2+2\Re\left(s_{14}\overline{s_{23}}\e^{\i(\theta_1+\theta_2)}\right)+
|s_{13}|^2+|s_{24}|^2+2\Re\left(s_{13}\overline{s_{24}}\e^{\i(\theta_1-\theta_2)}\right)\,.
\end{multline}
It obviously holds
\begin{gather*}
|s_{14}|^2+|s_{23}|^2+2\Re\left(s_{14}\overline{s_{23}}\e^{\i(\theta_1+\theta_2)}\right)\geq\left(|s_{14}|-|s_{23}|\right)^2\geq0\,,\\
|s_{13}|^2+|s_{24}|^2+2\Re\left(s_{13}\overline{s_{24}}\e^{\i(\theta_1-\theta_2)}\right)\geq\left(|s_{13}|-|s_{24}|\right)^2\geq0\,,
\end{gather*}
and $\theta_1,\theta_2$ can be chosen so that the terms on the second and third line of \eqref{p2q} are non-negative. Therefore
\begin{multline*}
\max\{p^2-q\,|\,\theta_1,\theta_2\in(-\pi,\pi]\}\\
\geq\frac{1}{4}\left(s_{11}+s_{22}-s_{33}-s_{44}\right)^2+\left(|s_{14}|-|s_{23}|\right)^2+\left(|s_{13}|-|s_{24}|\right)^2\geq0\,,
\end{multline*}
which we have set up to prove. \pfk

\begin{coro}
If $\max\{p^2-q\,|\,\theta_1,\theta_2\in(-\pi,\pi]\}=0$, then
$S=\mathrm{diag}(s_{11},s_{22},s_{33},s_{44})$ and
$s_{11}+s_{22}-s_{33}-s_{44}=0$.
\end{coro}
\pf It follows from the last part of the previous proof that the
premise requires $s_{11}+s_{22}-s_{33}-s_{44}=0$. If we substitute this into \emph{rhs} of \eqref{p2q} and make the same estimate as in the proof, we infer $s_{13}\overline{s_{23}}+s_{14}\overline{s_{24}}=0$, $\overline{s_{13}}s_{14}+\overline{s_{23}}s_{24}=0$. Then $\theta_1,\theta_2$ can be obviously chosen such that $p^2-q=\left(|s_{14}|+|s_{23}|\right)^2+\left(|s_{13}|+|s_{24}|\right)^2$, hence necessarily $s_{14}=s_{23}=s_{13}=s_{24}=0$. Now \eqref{p2q} gives $p^2-q=\left[\Re\left(s_{12}\e^{\i\theta_1}\right)-\Re\left(s_{34}\e^{\i\theta_2}\right)\right]^2$ which does not exceed zero only if $s_{12}=s_{34}=0$.
\pfk

\subsection{Point spectrum for a diagonal matrix $S$}

If $m=4$ and all the off-diagonal entries of $S$ vanish, then the
boundary conditions \eqref{bc:ST} describe the system of decoupled
edges of the length $a$. Such system has a point spectrum which is
not difficult to find. We have two edges types:
\begin{itemize}
 \item ``horizontal'', with the boundary conditions
$\psi'(0)=s_{22}\psi(0)$, $\psi'(a)=-s_{11}\psi(a)$,
 \item ``vertical'', with the boundary conditions $\psi'(0)=s_{44}\psi(0)$,
$\psi'(a)=-s_{33}\psi(a)$.
\end{itemize}
The corresponding spectral conditions are
 $$
 (s_{1+2j,1+2j}+ks_{2+2j,2+2j})\cos ka=(k-s_{1+2j,1+2j}s_{2+2j,2+2j})\sin
 ka\,, \quad j=0,1\,.
 $$
Their solutions are thus
 $$
 k=\frac{n\pi}{a}+\frac{\arctg s_{2+2j,2+2j}}{a}
 +\frac{s_{1+2j,1+2j}}{n\pi}+\O\left(n^{-2}\right)\,, \quad
 j=0,1\,,
 $$
giving rise the eigenvalues, or flat bands of the lattice
Hamiltonian,
 $$
 k^2=\left(\frac{n\pi}{a}\right)^2+2\,\frac{n\pi\,\arctg s_{2+2j,2+2j}}{a^2}
 +\left(\frac{\arctg s_{2+2j,2+2j}}{a}\right)^2
 +2\,\frac{s_{1+2j,1+2j}}{a}+\O\left(n^{-2}\right)
 $$
for $j=0,1$ corresponding the two edge orientations.

\subsection{A general matrix $S$}

Let us next consider the following sets related to solutions to
condition \eqref{d12},
$$
\mathcal{D}_{1,2}=\{-p\pm \sqrt{p^2-q}\,|\,\theta_1,\theta_2\in(-\pi,\pi]\}\cap\R\,,\\
$$
It follows from Proposition \ref{m4.p1} that
$\mathcal{D}_1\neq\emptyset\neq\mathcal{D}_2$. We will show now
that if one of them contains more than one point, then both
contain a non-degenerated interval.

\begin{prop}\label{m4.p2}
Let one of the expressions $-p+\sqrt{p^2-q}$, $-p-\sqrt{p^2-q}$ be
independent of $(\theta_1,\theta_2)$, then the same is true for
the other one.
\end{prop}
\pf Let us suppose that $-p+\sqrt{p^2-q}=c$ with $c\in\R$ holds
for $(\theta_1,\theta_2)$. In such a case $\sqrt{p^2-q}=p+c$,
hence $-q-2pc=c^2$, in other words, the expression $-q-2pc$ is
independent of $(\theta_1,\theta_2)$. Since
\begin{multline*}
-q-2pc=
\left(|s_{13}|^2-s_{11}s_{33}+|s_{14}|^2-s_{11}s_{44}+|s_{23}|^2-s_{22}s_{33}+|s_{24}|^2-s_{22}s_{44}\right)\\
+2\cdot(-1)^n\cdot\left[-(s_{33}+s_{44})\cdot\Re\left(s_{12}\e^{\i\theta_1}\right)+
\Re\left((s_{13}\overline{s_{23}}+s_{14}\overline{s_{24}})\e^{\i\theta_1}\right)\right.\\
\left.-(s_{11}+s_{22})\cdot\Re\left(s_{34}\e^{\i\theta_2}\right)+
\Re\left((\overline{s_{13}}s_{14}+\overline{s_{23}}s_{24})\e^{\i\theta_2}\right)\right]\\
+2\Re\left[(s_{14}\overline{s_{23}}-s_{12}s_{34})\e^{\i(\theta_1+\theta_2)}\right]
+2\Re\left[(s_{13}\overline{s_{24}}-s_{12}\overline{s_{34}})\e^{\i(\theta_1-\theta_2)}\right]\\
+c\cdot\left[s_{11}+s_{22}+s_{33}+s_{44}+2\cdot(-1)^n\cdot\Re\left(s_{12}\e^{\i\theta_1}\right)
+2\cdot(-1)^n\cdot\Re\left(s_{34}\e^{\i\theta_2}\right)\right]\,,
\end{multline*}
we easily find that also the expression
\begin{multline*}
(-1)^n\cdot\Re\left[\left(\left(-(s_{33}+s_{44})+c\right)\cdot s_{12}+
s_{13}\overline{s_{23}}+s_{14}\overline{s_{24}}\right)\e^{\i\theta_1}\right]\\
(-1)^n\cdot\Re\left[\left(\left(-(s_{11}+s_{22})+c\right)\cdot s_{34}+
\overline{s_{13}}s_{14}+\overline{s_{23}}s_{24}\right)\e^{\i\theta_2}\right]\\
+\Re\left[(s_{14}\overline{s_{23}}-s_{12}s_{34})\e^{\i(\theta_1+\theta_2)}\right]
+\Re\left[(s_{13}\overline{s_{24}}-s_{12}\overline{s_{34}})\e^{\i(\theta_1-\theta_2)}\right]
\end{multline*}
should be independent of
$(\theta_1,\theta_2)$. It follows from Proposition~\ref{LinNez}
that this is true if and only if the whole expression identically
equals zero. This in turn means that the \emph{lhs} of the
condition~\eqref{m4.4} is independent of $\theta_1,\theta_2$, and
consequently, both roots of the equation $d^2+2pd+q=0$ are
independent of $\theta_1,\theta_2$. \pfk

Proposition~\ref{m4.p2} in fact says that only two situations are
possible, either both sets $\mathcal{D}_1$, $\mathcal{D}_2$ are
non-degenerate intervals, or each of them contains a single
element only.

\begin{prop}
The sets $\mathcal{D}_1$, $\mathcal{D}_2$ are single-element sets
iff \emph{at most one} pair of off-diagonal elements of $S$ is
nonzero, and moreover, $s_{12}=s_{34}=0$.
\end{prop}
\pf $\mathcal{D}_1$, $\mathcal{D}_2$ are one-element \emph{iff}
both $-p+\sqrt{p^2-q}$, $-p-\sqrt{p^2-q}$ are independent of
$\theta_1,\theta_2$, and this obviously holds \emph{iff} both $p$,
$q$ are independent of $\theta_1,\theta_2$. With regard to the
definition of $p,q$, we have
\begin{gather*}
s_{12}=0\,,\quad s_{34}=0\,,\quad s_{13}\overline{s_{23}}+s_{14}\overline{s_{24}}=0\,,
\quad \overline{s_{13}}s_{14}+\overline{s_{23}}s_{24}=0\,,\\
s_{13}\overline{s_{24}}=0\,,\quad s_{14}\overline{s_{23}}=0\,.
\end{gather*}
Hence $s_{12}=0$, $s_{34}=0$, and it is easy to see that at least
three elements from the set $\{s_{13},s_{14},s_{23},s_{24}\}$ have
to vanish as well. \pfk

\begin{coro}
The sets $\mathcal{D}_1$, $\mathcal{D}_2$ are one-element sets
\emph{iff} the lattice decouples either into separate edges of the
length $a$, or into identical copies of L-shaped pairs of edges.
\end{coro}
\pf See the previous proposition. If all the off-diagonal elements
of $S$ vanish, the system decouples into separate edges and we
return the the situation considered above. The second possibility
is that just one of the numbers $s_{13}$, $s_{14}$, $s_{23}$,
$s_{24}$ is nonzero, then the system is decoupled into L-shaped
edge pairs. More precisely, $s_{13}\neq0$, $s_{14}\neq0$,
$s_{23}\neq0$ and $s_{23}\neq0$ corresponds to $\urcorner$,
$\lrcorner$, $\ulcorner$ and $\llcorner$ shaped pairs,
respectively, cf. Fig.~\ref{Mrizka} and the boundary
conditions~\eqref{vazba}. \pfk Apparently, if $s_{12}=s_{34}=0$
and $S$ contains just one nonzero off-diagonal element pair, the
spectrum consists of isolated points.

\subsection{Linearly growing spectral gaps, constant bands}

Consider now the case when at least two off-diagonal elements of
$S$ are nonzero, i.e. the situation when both $\mathcal{D}_1$,
$\mathcal{D}_2$ are intervals. Then the spectral asymptotics is
determined by~\eqref{m4.2} where $d$ is given by~\eqref{d12}.
Employing the symbols $\mathcal{D}_1,\mathcal{D}_2$ introduced at
the beginning of this section, we may characterize the values $k$
corresponding to the spectrum as
$$
k\in \bigcup_{j=1,2}
\left(\frac{n\pi}{a}+\frac{d_j^\downarrow}{n\pi}+\O(n^{-2}),\frac{n\pi}{a}+\frac{d_j^\uparrow}{n\pi}+\O(n^{-2})\right)\,,
$$
where $d_j^\downarrow=\min\mathcal{D}_j$, $d_j^\uparrow=\max\mathcal{D}_j$. Note
that $d_1^\pm,d_2^\pm$ depend only on $S$, since the term $(-1)^n$
can be absorbed into $\theta_1,\theta_2$. The above relation gives
spectral values in the form
\begin{equation} \label{m4.6}
k^2\in \bigcup_{j=1,2}
\left(\frac{n^2\pi^2}{a^2}+2\frac{d_j^\downarrow}{a}+\O(n^{-1}),\frac{n^2\pi^2}{a^2}+2\frac{d_j^\uparrow}{a}+\O(n^{-1})\right)\,,
\end{equation}
which means that spectral bands (or their overlapping pairs) are
of asymptotically constant width as the band number goes to
infinity, and consequently, the spectral gaps are linearly growing
with the band index. We also remark that the case a
\emph{nontrivial $\delta'_\mathrm{s}$ coupling} considered in
\cite{E1, EG} corresponds to $S$ with all the entries nonzero and
identical and thus has the described gap behaviour as expected.

The band structure depends on the parameter values, in particular,
the band corresponding to $\mathcal{D}_1$ and $\mathcal{D}_2$
in~\eqref{m4.6} may or may not overlap. The former situation
occurs, e.g., if only the elements $s_{12}$ and $s_{34}$ are
nonzero, the latter if $s_{11}+s_{22}-s_{33}-s_{44}$ is large in comparison
with the off-diagonal elements.


Let us finish this section with a short note on the prominent examples of the \emph{$\delta'_s$ and $\delta'$ coupling}.
A direct substitution of the appropriate boundary conditions into the expressions for $d_j$ leads to these results:
\begin{itemize}
\item In the case of the $\delta'_s$ coupling, it holds $d_2^\uparrow=d_1^\downarrow=0$. At the same time, one of the sets $\mathcal{D}_1,\mathcal{D}_2$ equals $\{d_2^\uparrow\}$. Consequently, the spectrum consists of asymptotically constant bands and linearly growing gaps between them. We remark that if $d_2^\uparrow=d_1^\downarrow$ and at the same time $\mathcal{D}_1$ and $\mathcal{D}_2$ are non-degenerate intervals, a special analysis is needed to find whether there is a gap between the bands corresponding to $\mathcal{D}_1$ and $\mathcal{D}_2$ or not.
\item In the case of the $\delta'$ coupling, we have $d_2^\uparrow<d_1^\downarrow$, thus the sets
$\mathcal{D}_1$ and $\mathcal{D}_2$ are disjoint. Consequently, the spectrum has
asymptotically the pattern $\cdots GbgbGbgbGbgbG\cdots$ where $G$
represents linearly growing gaps and $b,g$ stand for bands and
gaps whose widths are asymptotically constant.
\end{itemize}


\section*{Appendix: proof of Proposition~\ref{m2.prop}}

We will prove the claims in the order of their presentation: \\
[.5em]
\textit{(i)} It holds $-K_c\cos^2 x+K_s\sin^2
x=\frac{-K_c+K_s}{2}-\frac{K_c+K_s}{2}\cos 2x$ and
\begin{multline*}
\max\left\{\left.\cos x\cdot L_c(\theta_1,\theta_2)+L(\theta_1,\theta_2)\,\right|\,\theta_1,\theta_2\in(-\pi,\pi]\right\}\\
=\max\left\{\left.\cos x\cdot L_c(\theta_1,\theta_2)+L(\theta_1,\theta_2)\,\right|\,\theta_1,\theta_2\in(-2\pi,0]\right\}\\
=\max\left\{\left.\cos x\cdot L_c(\theta_1-\pi,\theta_2-\pi)+L(\theta_1-\pi,\theta_2-\pi)\,\right|\,\theta_1,\theta_2\in(-\pi,\pi]\right\}\\
=\max\left\{\left.-\cos x\cdot
L_c(\theta_1,\theta_2)+L(\theta_1,\theta_2)\,\right|\,\theta_1,\theta_2\in(-\pi,\pi]\right\}\,,
\end{multline*}
therefore $V_2^+(x)$ equals
$$
\frac{-K_c+K_s}{2}-\frac{K_c+K_s}{2}\cos 2x
+\max\left\{\left.|\cos x|\cdot L_c(\theta_1,\theta_2)
+L(\theta_1,\theta_2)\,\right|\,\theta_1,\theta_2\in(-\pi,\pi]\right\}\,,
$$
and the function $V_2^+(x)$ can be treated in the same way. To
finish the proof of \textit{(i)}, it suffices to realize that both
$\cos 2x$ and $|\cos x|$ are $\pi$-periodic functions satisfying
$\cos 2\left(\frac{\pi}{2}-x\right)=\cos 2x$ and
$|\cos\left(\frac{\pi}{2}-x\right)|=|\cos x|$.

\textit{(ii)} The part of $V_2^-(x)$ independent of
$\theta_1,\theta_2$ is obviously increasing on
$\left[0,\frac{\pi}{2}\right]$, because $-\cos 2x$ does. As for
the second part, for all $x'\in\left[0,\frac{\pi}{2}\right)$ it
holds: if $\tilde{\theta_1},\tilde{\theta_2}$ satisfy
$$
\min\left\{\left.\cos x'\cdot L_c(\theta_1,\theta_2)
+L(\theta_1,\theta_2)\,\right|\,\theta_1,\theta_2\in(-\pi,\pi]\right\}
=\cos x'\cdot
L_c(\tilde{\theta_1},\tilde{\theta_2})+L(\tilde{\theta_1},\tilde{\theta_2})\,,
$$
then $L_c(\tilde{\theta_1},\tilde{\theta_2})\leq0$. If this was
not the case one could shift both
$\tilde{\theta_1},\tilde{\theta_2}$ by $\pi$, which would change
the sign of $L(\tilde{\theta_1},\tilde{\theta_2})$, and therefore
make the value of $\cos x'\cdot
L_c(\tilde{\theta_1},\tilde{\theta_2})+L(\tilde{\theta_1},\tilde{\theta_2})$
smaller, however, this would contradict the assumed minimality.

Let now $x''<x'\in\left(0,\frac{\pi}{2}\right]$. Then $\cos
x''\cdot
L_c(\tilde{\theta_1},\tilde{\theta_2})+L(\tilde{\theta_1},
\tilde{\theta_2})\leq\cos x'\cdot
L_c(\tilde{\theta_1},\tilde{\theta_2})
+L(\tilde{\theta_1},\tilde{\theta_2})$, and therefore
\begin{multline*}
\min\left\{\left.\cos x''\cdot
L_c(\theta_1,\theta_2)+L(\theta_1,\theta_2)\,
\right|\,\theta_1,\theta_2\in(-\pi,\pi]\right\}\\
\leq\min\left\{\left.\cos x'\cdot
L_c(\theta_1,\theta_2)+L(\theta_1,\theta_2)\,
\right|\,\theta_1,\theta_2\in(-\pi,\pi]\right\}\,,
\end{multline*}
thus $\min\left\{\left.\cos x'\cdot L_c(\theta_1,\theta_2)
+L(\theta_1,\theta_2)\,\right|\,\theta_1,\theta_2\in(-\pi,\pi]\right\}$
is an increasing function of $x$ in the interval
$\left[0,\frac{\pi}{2}\right]$.

\textit{(iii)} Let $\tilde{\theta_1}(x)$ and $\tilde{\theta_2}(x)$
be functions defined on $\left[0,\frac{\pi}{2}\right]$ such that
\begin{eqnarray*}
\lefteqn{\max\left\{\left.\cos x\cdot L_c(\theta_1,\theta_2)
+L(\theta_1,\theta_2)\,\right|\,\theta_1,\theta_2\in(-\pi,\pi]\right\}}
\\ && =\cos x\cdot
L_c(\tilde{\theta_1}(x),\tilde{\theta_2}(x))+L(\tilde{\theta_1}(x),\tilde{\theta_2}(x))\,;
\end{eqnarray*}
it follows that
\begin{equation}\label{parcialni}
\cos x\cdot\frac{\partial
L_c}{\partial\tilde{\theta_j}}(\tilde{\theta_1}(x),\tilde{\theta_2}(x))
+\frac{\partial
L}{\partial\tilde{\theta_j}}(\tilde{\theta_1}(x),\tilde{\theta_2}(x))=0\,,
\qquad j=1,2,
\end{equation}
which we will use below. The function $V_2^+(x)$ can be then
written as
$$
V_2^+(x)=\frac{-K_c+K_s}{2}-\frac{K_c+K_s}{2}\cos 2x +\cos x\cdot
L_c(\tilde{\theta_1}(x),\tilde{\theta_2}(x))+L(\tilde{\theta_1}(x),\tilde{\theta_2}(x))\,,
$$
so we can compute its derivative with respect to $x$,
\begin{multline*}
\frac{\d}{\d x} V_2^+(x)=(K_c+K_s)\sin2x
-\sin x\cdot L_c(\tilde{\theta_1}(x),\tilde{\theta_2}(x))\\
+\cos x\left(\frac{\partial L_c}{\partial \tilde{\theta_1}}
\tilde{\theta_1}'(x) +\frac{\partial L_c}{\partial
\tilde{\theta_2}}\tilde{\theta_2}'(x)\right)+ \frac{\partial
L}{\partial \tilde{\theta_1}}\tilde{\theta_1}'(x) +\frac{\partial
L}{\partial \tilde{\theta_2}}\tilde{\theta_2}'(x)\,.
\end{multline*}
The expression on the second line vanishes due
to~\eqref{parcialni}, hence
$$
\frac{\d}{\d x} V_2^+(x)=\sin x\left[2(K_c+K_s)\cos x
-L_c(\tilde{\theta_1}(x),\tilde{\theta_2}(x))\right]\,.
$$
Now we will show that $\frac{\d}{\d x} V_2^+(x)\leq0\Rightarrow
V_2^+(x)\geq1$ --- once this is proved, the proof of statement
\textit{(iii)} is finished. We have $\frac{\d}{\d x}
V_2^+(x)\leq0\Rightarrow -K_c\cos x\geq K_s\cos
x-\frac{1}{2}L_c(\tilde{\theta_1}(x),\tilde{\theta_2}(x))$. A
substitution for $K_c\cos x$ into $V_2^+(x)$ gives the inequality
$$
V_2^+(x)\geq K_s+\frac{1}{2}\cos x\cdot L_c(\tilde{\theta_1}(x),
\tilde{\theta_2}(x))+L(\tilde{\theta_1}(x),\tilde{\theta_2}(x))\,.
$$
Since $K_s\geq1$, it only remains to check that $\frac{1}{2}\cos
x\cdot L_c(\tilde{\theta_1}(x), \tilde{\theta_2}(x))
+L(\tilde{\theta_1}(x), \tilde{\theta_2}(x))\geq0$. The argument
leans on the following two statements:
\begin{itemize}
\item[\textit{S1.}]
$L_c(\tilde{\theta_1}(x),\tilde{\theta_2}(x))\geq0$.
\item[\textit{S2.}] If
$L(\tilde{\theta_1}(x),\tilde{\theta_2}(x))<0$, then $\cos x\cdot
L_c(\tilde{\theta_1}(x),\tilde{\theta_2}(x))\geq
2|L(\tilde{\theta_1}(x),\tilde{\theta_2}(x))|$.
\end{itemize}
Statement \textit{S1} can be demonstrated using the same idea as
in part \emph{(ii)} of this proof, \textit{S2} can be proved by
\emph{reductio ad absurdum}. Let us suppose that
$L(\tilde{\theta_1}(x),\tilde{\theta_2}(x))<0$ and $|\cos x\cdot
L_c(\tilde{\theta_1}(x),\tilde{\theta_2}(x))|<
2|L(\tilde{\theta_1}(x),\tilde{\theta_2}(x))|$. Then we define
$\hat{\theta_1}=\tilde{\theta_1}(x)+\rho$ and
$\hat{\theta_2}=\tilde{\theta_2}(x)+\rho$ where
$\rho=\frac{\pi}{2}$ or $\rho=-\frac{\pi}{2}$ --- the sign of
$\rho$ is chosen such that
$L_c(\hat{\theta_1},\hat{\theta_2})\geq0$. Then it holds
\begin{multline*}
\cos x\cdot L_c(\hat{\theta_1},\hat{\theta_2})
+L(\hat{\theta_1},\hat{\theta_2})\geq \cos x\cdot
0-L(\tilde{\theta_1}(x),\tilde{\theta_2}(x))
=|L(\tilde{\theta_1}(x),\tilde{\theta_2}(x))|\\
>\frac{1}{2}\cos x\cdot L_c(\tilde{\theta_1}(x),\tilde{\theta_2}(x))>
\frac{1}{2}\cos x\cdot L_c(\tilde{\theta_1}(x),
\tilde{\theta_2}(x))+L(\tilde{\theta_1}(x),\tilde{\theta_2}(x))\,,
\end{multline*}
hence
$$
V_2(x,\hat{\theta_1},\hat{\theta_2})>V_2(x,\tilde{\theta_1}(x),\tilde{\theta_2}(x))=V_2^+(x)\,,
$$
which is in contradiction with the definition of $V_2^+(x)$.

\textit{(iv)} It follows from their construction that the
functions $V_2^+$, $V_2^-(x)$ satisfy $V_2^+(x)\geq V_2^-(x)$ for
all $x\in\left(0,\frac{\pi}{2}\right)$. The strict inequality
$V_2^+(x)>V_2^-(x)$ is equivalent to the fact that the image of
the function $\cos x\cdot L_c(\theta_1,\theta_2)
+L(\theta_1,\theta_2)$ defined on $(-\pi,\pi]^2$ forms a
non-degenerate interval. Since $\cos x\cdot
L_c(\theta_1,\theta_2)+L(\theta_1,\theta_2)$ equals
\begin{multline*}
\Re\left(-8\cos
ak(t_{11}\overline{t_{21}}+t_{12}\overline{t_{22}})\e^{\i\theta_1}\right)+
\Re\left(8\cos ak(t_{22}\overline{t_{21}}+\overline{t_{11}}t_{12})\e^{\i\theta_2}\right)\\
+\Re\left(8t_{11}\overline{t_{22}}\e^{\i(\theta_1-\theta_2)}\right)
+\Re\left(8t_{12}\overline{t_{21}}\e^{\i(\theta_1+\theta_2)}\right)\,,
\end{multline*}
i.e. it is of the type examined in Proposition~\ref{LinNez}, we
infer that the image degenerates to a single point \emph{iff}
$$
t_{11}\overline{t_{21}}+t_{12}\overline{t_{22}}=0 \quad\wedge\quad
t_{22}\overline{t_{21}}+\overline{t_{11}}t_{12}=0 \quad\wedge\quad
t_{11}\overline{t_{22}}=0 \quad\wedge\quad
t_{12}\overline{t_{21}}=0\,;
$$
it is straightforward to check that these four conditions are
satisfied \emph{iff} at most one of the numbers
$t_{11},t_{12},t_{21},t_{22}$ is nonzero. This concludes the
proof.

\section*{Acknowledgments}
We dedicate this paper to memory of Pierre Duclos, a longtime
friend and coauthor of the first author and a thesis supervisor of
the second one. The research was supported by the Czech Ministry
of Education, Youth and Sports within the project LC06002. We
thank the referees for comments which helped to improve the text.

\end{document}